\documentclass[a4paper,11pt]{article}
\pdfoutput=1
\usepackage{jheppub}
\usepackage[T1]{fontenc}
\usepackage{physics}
\usepackage{amsmath}
\usepackage{amsthm}
\usepackage{amsfonts}
\usepackage{amssymb}
\usepackage{graphicx, rotating}
\usepackage{epstopdf}
\usepackage{epsfig}
\usepackage{latexsym}
\usepackage{graphicx}
\usepackage{color}
\usepackage{slashed}
\usepackage{float}
\usepackage[export]{adjustbox}
\usepackage{simplewick}
\usepackage{subcaption}

\usepackage{textgreek}
\usepackage[dvipsnames]{xcolor}
\definecolor{rossos}{cmyk}{0,1,1,0.55}
\definecolor{bluscuro}{rgb}{0.15, 0.2, .85}
\definecolor{bluchiaro}{cmyk}{1,.3,0.,0.1}

\newcommand{\eq}[1]{Eq.~(\ref{#1})}

\newcommand{\nn}{\nonumber}

\newcommand{\be}{\begin{equation}}
\newcommand{\ee}{\end{equation}}
\newcommand{\bea}{\begin{eqnarray}}
\newcommand{\eea}{\end{eqnarray}}
\newcommand{\bc}{\begin{center}}
\newcommand{\ec}{\end{center}}

\DeclareMathOperator{\diag}{diag}

\def\ee{\text{e}}

\usepackage{graphicx,wrapfig}

\usepackage{amsthm,bm}
\usepackage{gensymb}

\usepackage{array}
\usepackage{tcolorbox, mathtools}
\usepackage{soul}
\usepackage{feynmp-auto}

\newcommand{\kf}[1]{\textcolor{violet}{[KF: #1]}}

\usepackage{lmodern}
\usepackage[T2A,T1]{fontenc}
\usepackage[utf8]{inputenc}
\usepackage[utf8]{inputenc}
\usepackage{bm}
\usepackage{mathtools}
\usepackage{amsmath}
\usepackage{slashed}	
\usepackage{bbm}

\usepackage{tikz}
\usetikzlibrary{snakes}
\usetikzlibrary{decorations}
\usetikzlibrary{trees}
\usetikzlibrary{decorations.pathmorphing}
\usetikzlibrary{decorations.markings}
\usetikzlibrary{external}
\usetikzlibrary{intersections}
\usetikzlibrary{shapes,arrows}
\usetikzlibrary{arrows.meta}
\usetikzlibrary{calc}
\usetikzlibrary{shapes.misc}
\usetikzlibrary{decorations.text}
\usetikzlibrary{backgrounds}
\usetikzlibrary{fadings}
\usetikzlibrary{tikzmark,calc,arrows,shapes,decorations.pathreplacing}

\tikzset{
	graviton/.style={decorate,line width=0.15mm, 
	decoration={snake,amplitude=.6mm, segment length=1.5mm}
	},
	photon/.style={decorate, decoration={snake}, draw=red},
	scalar/.style={postaction={decorate},
	},
	massive/.style={postaction={decorate},
		line width=0.75mm,
	},
	massless/.style={postaction={decorate},
	},
	masslessWithDot/.style={postaction={decorate},
		decoration={
			markings,
			mark=at position 0.5 with {\fill circle (2pt);}}
	},
	massiveWithDot/.style={postaction={decorate},
		line width=0.5mm,
		decoration={
			markings,
			mark=at position 0.5 with {\fill circle (2pt);}}
	},
	massiveWithArrow/.style={postaction={decorate},
		line width=0.75mm,
		decoration={
			markings,
			mark=at position 0.5 with {\arrow{latex}}}
	},
	massiveWithArrowB/.style={postaction={decorate},
		line width=0.75mm,
		decoration={
			markings,
			mark=at position 0.95 with {\arrow{latex}}}
	},
	fermion/.style={postaction={decorate},
		line width=0.4mm,
		decoration={
			markings,
			mark=at position 0.5 with {\arrow{latex[reversed]}}}
    },
	massiveLin/.style={postaction={decorate},
		double,
		thick,
		fill=white
	},
	massivePhi/.style={postaction={decorate},
		line width=0.75mm,
		dashed
	},
	masslessPhi/.style={postaction={decorate},
		dashed
	},
	unitaryCut/.style={postaction={draw,densely dashed,blue,thin},
		line width = 0.2cm,white
	},
	gluon/.style={decorate, draw=magenta,
		decoration={coil,amplitude=4pt, segment length=5pt}},
	partial ellipse/.style args={#1:#2:#3}{
		insert path={+ (#1:#3) arc (#1:#2:#3)}
	},
	cross/.style={cross out, draw=black, minimum size=2*(#1-\pgflinewidth), inner sep=0pt, outer sep=0pt},
	branchCut/.style={postaction={decorate},
		snake=zigzag,
		decoration = {snake=zigzag,segment length = 2mm, amplitude = 2mm}	
	}
	cross/.default={1pt}
}

\colorlet{mred}{black!30!red}
\colorlet{mgreen}{black!30!green}
\colorlet{mblue}{black!30!blue}
\colorlet{morange}{blue!70!red}

\graphicspath{{figures/}}
\usepackage[export]{adjustbox}

\def\nn{\nonumber}

\newcommand{\figref}[1]{Fig.~\ref{#1}}

\def\psint2{\mathcal{I}_2} 
\def\psintN2{\hat{\mathcal{I}}_2} 

\begin{document}

\title{Bound States in 2d Yukawa Theory from Hamiltonian Truncation}

\author[a]{Olivier Delouche,}
\emailAdd{olivier.delouche@unige.ch}

\author[a,b]{Kara Farnsworth,}
\emailAdd{kmfarnsworth@gmail.com}

\author[a]{Francesco Riva,}
\emailAdd{francesco.riva@unige.ch}

\affiliation[a]{D\'epartement de Physique Th\'eorique, Universit\'e de Gen\`eve,
24 quai Ernest-Ansermet, 1211 Gen\`eve 4, Switzerland}
\affiliation[b]{Maxwell Institute for Mathematical Sciences, Department of Mathematics,
Heriot-Watt University, Edinburgh EH14, UK}

\abstract{
We study the spectrum of two-dimensional Yukawa theory using Hamiltonian truncation, with a focus on  bound states. We extend the framework of Hamiltonian truncation effective theory to include fermions and logarithmic UV divergences. With these corrections, the energy levels converge as a function of the cutoff with the scaling predicted by power counting. We identify a strategy to follow the evolution of fermion-antifermion bound states from weak to strong coupling, and investigate its binding energy, finding good agreement with finite-volume perturbation theory at weak coupling. 
Finally, we explore the heavy-scalar regime toward the massive Thirring/Sine-Gordon limit, recovering the expected weak-coupling behavior and outlining how the calculation can be extended toward this limit.
}

\maketitle

\section{Motivation}
\label{sec:intro}
Strongly coupled quantum field theories (QFTs) describe many of the most intriguing phenomena across theoretical physics, ranging from the strong nuclear force to strongly correlated condensed matter systems. Although these theories model physically important systems, their behavior remains extremely difficult to predict  when perturbation theory breaks down and qualitatively new effects can emerge. Yukawa theory provides a relatively simple setting for studying this regime with both bosonic and fermionic degrees of freedom. Yukawa interactions, and their generalizations, provide a mechanism for generating fermion masses through symmetry breaking, most notably through the coupling of Standard Model fermions  -- in particular the top quark -- to the Higgs field. The same interactions generate attractive forces through boson exchange, producing bound states and qualitatively changing the spectrum. Yukawa theories can also exhibit critical behavior and nontrivial high-energy scattering such as Regge behavior \cite{Cresswell-Hogg:2023rvu, Ahmadzadeh:1963ith}. In this paper, we study Yukawa theory in $1+1$ dimensions, where analytic results are available in several limits, and investigate its strong-coupling
regime.

To make reliable predictions for this and other systems at strong coupling, we require genuinely nonperturbative techniques. Methods like lattice field theory \cite{Gattringer:2010zz} and the numerical bootstrap \cite{Poland:2018epd} have had great success in studies of strongly coupled phenomena, but each has its limitations. A complementary nonperturbative method is Hamiltonian truncation. As a continuum Hamiltonian method, it has direct access to real-time observables, and since it avoids a spatial lattice it can study theories that are difficult to discretize, such as those with chiral fermions or supersymmetry \cite{Nielsen:1980rz}. Unlike the numerical bootstrap, which generally requires substantial additional input to isolate a specific theory, Hamiltonian truncation can probe a particular theory and directly calculate its spectrum and dynamics. The method was first developed for two-dimensional QFTs \cite{Brooks:1983sb}, including discretized lightcone quantization \cite{Pauli:1985ps,Brodsky:1997de} and the truncated conformal space approach \cite{Yurov:1989yu,Yurov:1991my}. Interest in Hamiltonian truncation as a more general method was recently renewed \cite{Katz:2013qua,Katz:2014uoa,Hogervorst:2014rta,Rychkov:2014eea}, leading to applications across a wider range of theories and observables \cite{Rychkov:2015vap,Lencses:2015bpa, Bajnok:2015bgw,Rakovszky:2016ugs,Anand:2017yij,Hogervorst:2018otc,Delacretaz:2018xbn,Fitzpatrick:2019cif,Elias-Miro:2020qwz,Anand:2020qnp,Hogervorst:2021spa,Chen:2021bmm,Anand:2021qnd,Delacretaz:2022ojg,Henning:2022xlj,Fitzpatrick:2023mbt,Fitzpatrick:2023aqm,Fitzpatrick:2024rks,Ingoldby:2025bdb,Fitzpatrick:2025hqk,Houtz:2025lbv} and the development of new techniques to refine and extend the method \cite{Katz:2016hxp,EliasMiro:2021aof,Emonts:2022vim,Chen:2022zms,Schmoll:2023eez,Ingoldby:2024fcy,Basak:2026meu,Elias-Miro:2015bqk,Elias-Miro:2017tup,Rutter:2018aog,Fitzpatrick:2018ttk,Cohen:2021erm,EliasMiro:2022pua,Chen:2023glf,Lajer:2023unt,Delouche:2023wsl,Delouche:2024yuo,Demiray:2025zqh,Maestri:2026hqb,Li:2026dyb,Houtz:2026wra} (see \cite{James:2017cpc, Anand:2020gnn,Fitzpatrick:2022dwq} for reviews).

Hamiltonian truncation is a variational method that separates the Hamiltonian $H$ of a system into a solvable part $H_0$ and an interaction $V$,
\begin{align}
H = H_0 + V.
\end{align}
Contrary to perturbation theory, $V$ need not be small in the conventional sense. The full Hamiltonian is then diagonalized using a finite set of $H_0$ eigenstates with energy below a cutoff $E_{\rm max}$. For sufficiently relevant interactions $V$, the effects of the omitted high-energy states are increasingly suppressed as $E_{\max}$ is raised, and the method converges quickly with increasing cutoff. However, larger $E_{\max}$ also leads to an
exponential growth in the size of the truncated Hilbert space, making theories with many degrees of freedom computationally intractable. Because of this, Hamiltonian truncation has historically been restricted to relatively simple systems, and much of its potential  remains unexplored.

In order to overcome this exponential barrier, several methods have been developed to incorporate the effects of the states omitted by  truncation \cite{Feverati:2006ni, Giokas:2011ix, Hogervorst:2014rta, Rychkov:2014eea, Lencses:2015bpa,Elias-Miro:2015bqk, Elias-Miro:2017tup, Rutter:2018aog, Fitzpatrick:2018ttk, Elias-Miro:2020qwz, Anand:2020gnn,  Cohen:2021erm, EliasMiro:2022pua, Chen:2023glf, Lajer:2023unt, Delouche:2023wsl,Delouche:2024yuo,Demiray:2025zqh,Maestri:2026hqb}. These techniques add corrections to the truncated Hamiltonian, with the goal of improving accuracy without increasing the size of the basis. Although the different approaches for calculating these corrections are similar in spirit, the one we use here is guided by the logic of effective field theory (EFT). EFT gives a systematic prescription for incorporating the effects of high-energy physics into a simpler description valid at low energies. Adapting this idea to Hamiltonian truncation gives Hamiltonian truncation effective theory (HTET), which has been shown to improve convergence in two-dimensional scalar theories \cite{Cohen:2021erm} and can be systematically improved by the inclusion of nonlocal  corrections \cite{Demiray:2025zqh}. Although these initial
results are promising, this method must be tested in more contexts to
understand its versatility and identify potential shortcomings.

The two-dimensional Yukawa model provides a natural next step for testing this formalism, while also exhibiting interesting strongly-coupled behavior of its own. This model was studied in some of the earliest applications of Hamiltonian truncation \cite{Brooks:1983sb,Pauli:1985ps}. More recently lightcone conformal truncation was used to study  Yukawa theories with Majorana fermions \cite{Fitzpatrick:2019cif,Anand:2020gnn}, but the strong-coupling behavior of the Dirac Yukawa model remains less well understood. In this work we revisit this theory using equal-time Hamiltonian truncation together with EFT corrections computed using HTET. 

Besides its physical interest, this model contains several features that make it a useful test of the wider applicability of this formalism. In two dimensions, the Yukawa model has nontrivial UV divergences that cannot be removed by  normal ordering. Instead, the theory must be renormalized by adding the appropriate counterterms before matching onto the truncated Hamiltonian. Another complication is that the distinction between bound and scattering states becomes unambiguous only in the infinite-volume limit~\cite{Luscher:1986pf,Luscher:1985dn}, while our numerical calculations are necessarily performed at finite volume. Showing how to overcome these complications is one of the goals of this work. In the regime where the scalar is heavy, integrating it out generates at leading order a four-fermion interaction, which is \emph{marginal} in two dimensions. The convergence of Hamiltonian truncation for marginal interactions remains poorly understood~\cite{Beria:2013hz,Rutter:2018aog}.
The Yukawa model provides a useful alternative in which the marginal interaction emerges at low energies from a high energy theory containing only relevant couplings. Understanding how Hamiltonian truncation reproduces this separation of scales will be important for extending the method to more realistic QFTs, including four-dimensional gauge theories such as QCD \cite{Fitzpatrick:2022dwq}.

The rest of this paper is organized as follows. In Sec.~\ref{sec:an} we review
known results for bound states in Yukawa theory, including the weak-coupling
non-relativistic limit, the large-$m_\phi$ Thirring/Sine-Gordon limit, and
finite-volume effects. In Sec.~\ref{sec:HT} we construct the truncated Yukawa
Hamiltonian and derive the counterterms and effective Hamiltonian corrections
using HTET. In Sec.~\ref{sec:results} we study the convergence of the corrected
Hamiltonian and present results for the low-lying spectrum and bound states. We conclude in Sec.~\ref{sec:conclusions}.

\section{Known results in Yukawa theory}
\label{sec:an}
The theory we are interested in is the Yukawa model in 1+1 dimensions,
\begin{align}\label{eqn:Lyuk}
\mathcal{L} = 
 \frac{1}{2} \partial_\mu \phi \partial^\mu \phi -\frac{1}{2} m_\phi^2 \phi^2+ \bar{\psi}\left(i \slashed{\partial} - m_\psi\right) \psi - g \phi \bar{\psi} \psi 
\end{align}
where $\psi$ is a two-component (Dirac) fermion.
Besides the $U(1)$ symmetry acting on the fermion, this theory is invariant under both charge conjugation $C$ and parity, and has a $\mathbb{Z}_2$ symmetry ($\phi\to-\phi$, $\psi\to \gamma_* \psi$) broken by $m_\psi$.
\enlargethispage{2\baselineskip}
\footnote{We choose our basis of $\gamma$ matrices to be
\begin{align}
\gamma^0 = \begin{pmatrix} 0 & 1 \\
1 &0 
\end{pmatrix}, \quad{} \gamma^1 = \begin{pmatrix}
0 & 1\\
-1 & 0
\end{pmatrix}
, \quad{} \gamma_* = \begin{pmatrix}
-1 & 0\\
0 & 1
\end{pmatrix},
\end{align}
where $\gamma_* = \gamma^0 \gamma^1$ is the equivalent of the four-dimensional $\gamma_5$.
}

We want to understand the spectrum of this theory, both at infinite and finite volume $L$, in order to compare with the Hamiltonian truncation results in Sec.~\ref{sec:results}.
This theory has at least one bound state below threshold, and in certain regimes (part of) the spectrum is known analytically. We summarize these results below.

\subsection{Non-relativistic limit}\label{sec:NRPTs}
In infinite volume $L\to\infty$ and at weak coupling  $g/m_\phi\ll 1$ and $g/m_\psi\ll 1$, the spectrum develops a fermion-antifermion bound state  governed by non-relativistic dynamics.
It can be studied using the Schr\"odinger equation
with Yukawa potential $V=-\,\frac{g^2}{2m_\phi}\,e^{-m_\phi|x|}$,
\begin{equation}
-\frac{1}{m_\psi}\,\Psi''(x) - \frac{g^2}{2m_\phi}\,e^{-m_\phi|x|}\,\Psi(x)= -E_B\,\Psi(x)
\label{eq:Schr}
\end{equation}
where $E_B>0$ is the binding energy and $\Psi$ is the bound-state wavefunction. 
For $x>0$, with a change of coordinates this becomes Bessel's equation. The solution that is regular as $x\to\infty$ is
\begin{equation}
 \Psi(x)=N\,J_\nu\!\,\big(\alpha e^{-m_\phi x/2}\big)\qquad \textrm{with}\qquad \nu\equiv\frac{2\sqrt{m_\psi E_B}}{m_\phi}\,,\qquad  \alpha \equiv\frac{2g}{m_\phi}\sqrt{\frac{m_\psi}{2m_\phi}}
\label{eq:phys-solution}
\end{equation}
where $N$ is a normalization constant. For $x<0$ we can repeat these steps and obtain the same form since $V$ is even under parity $x\to-x$. 
Gluing the $x\lessgtr0$ wavefunctions together we have parity-even (odd)  solutions, with boundary conditions $J'_\nu (\alpha)=0$ ($J_\nu (\alpha)=0$). These impose conditions on $\nu$, which determine the binding energies and the number of bound states for a given $\alpha$.

  At small $\alpha$ there is always a parity-even bound state. Using   $0=J_\nu'(\alpha)=\frac{\nu}{\alpha}J_\nu(\alpha)-J_{\nu+1}(\alpha)$ and the small $\alpha$-expansion $J_\nu(\alpha)\approx (\alpha/2)^\nu/\Gamma(1+\nu)$, we find,
\begin{equation}\label{exres}
\nu\simeq \alpha^2/2 \quad\Rightarrow  \quad E_B\simeq \frac{g^4 m_\psi}{4m_\phi^4}+O\left(\frac{g^6m_\psi^2}{m_\phi^7}\right),
\end{equation}
which, importantly, lacks an $O(g^2)$ term.
In this regime, the fermion-antifermion bound-state system with reduced mass $m_\psi/2$ has typical relative velocity and momentum,
\begin{equation}\label{limitsnr}
v_B\approx \sqrt{\frac{4 E_B}{m_\psi}}\simeq  \frac{g^2}{m_\phi^2}\,,\qquad \kappa_B\equiv \sqrt{m_\psi E_B}\simeq m_\psi \frac{ g^2}{2m_\phi^2}\,.
\end{equation}
At weak coupling $g/m_\phi\ll1$, the bound state is non-relativistic, and we can use the momentum to associate a size $L_B\equiv\kappa_B^{-1}$ to the bound state.

Parity-odd solutions appear at $\alpha\gtrsim \bar \alpha \equiv 2.4$ (the location of the first zero of $J_0(\alpha)$), which requires either strong coupling or a large hierarchy $m_\psi/m_\phi$.
Their binding energy can be found by expanding the boundary condition close to this zero,
\begin{equation}
     E_B^{\rm odd} \simeq 0.2 m_\phi \left(\frac{g}{m_\phi}-\frac{\bar g}{m_\phi}\right)^2,\qquad \frac{\bar g}{m_\phi} \equiv \bar \alpha\sqrt\frac{m_\phi}{{2 m_\psi}}\,.\label{eq:secondBS}
\end{equation}
Other bound-state solutions require even larger couplings or mass hierarchies, but as long as they fall within the non-relativistic approximation they can be found from the zeros of $J_\nu, J_\nu^\prime$ as done here.

\subsection{Large $m_\phi$ limit}
Outside the non-relativistic regime, no general analytic solution is known for Yukawa theory. However, an analytically tractable limit exists when $g,\,m_\phi\to\infty$ with $g/m_\phi$ fixed.
In this limit, integrating out the scalar gives at leading $1/m_\phi$ order a four-fermion interaction that can be written as the massive
Thirring model~\cite{Thirring:1958in},\footnote{In the form obtained by integrating out the scalar, this is the
$N=1$ Gross-Neveu model~\cite{Gross:1974jv}, which is equivalent to the
massive Thirring model via the two-dimensional Fierz identity
$(\bar\psi\gamma_\mu\psi)^2=-2(\bar\psi\psi)^2$. Higher orders in the EFT expansion are suppressed by powers of $m_\phi$ but do not receive any $g$-enhancement.}
\begin{align}
\mathcal{L}_{\rm Th} 
&= \bar{\psi}\left(i \slashed{\partial} - m_\psi\right) \psi -\frac{g^2}{4m_\phi^2}( \bar{\psi}\gamma_\mu  \psi)^2. 
\end{align}

This theory is exactly dual to the Sine-Gordon model~\cite{Mandelstam:1975hb,Coleman:1974bu},
\begin{equation}
\mathcal L_{\rm SG}=\frac{1}{2}(\partial_\nu\varphi)(\partial^\nu\varphi)+\mu\,\cos(\beta\varphi),
\end{equation}
with the identification 
\begin{equation}
{\,\frac{4\pi}{\beta^2}=1+\frac{g^2}{2\pi m_\phi^2}\,},
\end{equation}
where $\mu$ is a function of $m_\psi$ and $g/m_\phi$  that we will not need explicitly.
In infinite volume, the Sine-Gordon spectrum  is known exactly and consists of a soliton and antisoliton -- which are dual to the Thirring fermion and antifermion -- as well as breathers, which are bound states in the fermionic theory. The soliton mass $M_s$ is identified with the physical fermion mass $M_s=m_{\psi,\rm phys}$. The breather masses are
\begin{equation}\label{eq:breathers}
{M_{n}=2 M_s\,\sin\!\Big(\frac{n\pi\xi}{2}\Big),\quad n=1,\dots,N_b\, } 
\end{equation}
where $\xi^{-1} \equiv 1+g^2/(\pi m_\phi^2)$ and $N_b=\lfloor {\xi^{-1}}\rfloor$. With the binding energy of the first breather defined as $E_B\equiv 2M_s-M_1$, the typical momenta/inverse size of the bound state are~$\kappa_B=\sqrt{M_s E_B - E_B^2/4}=L_B^{-1}$.

For $g^2/m_\phi^2<\pi$ the theory supports a unique bound state,\footnote{Contrary to the finite $m_\phi$ case, the second bound state is always in the relativistic/nonperturbative regime.}
and at weak coupling $g/m_\phi\ll 1$  its binding energy is
\begin{equation}\label{Ebapp}
   E_B=M_s \left( \frac{g^4}{4 {m_\phi}^4}-\frac{g^6}{2\pi  {m_\phi}^6}+O\left(\frac{g^8}{m_\phi^8}\right)\right),
\end{equation}
which starts at order $g^4$, as expected from the perturbative result. A second bound state appears for $g^2/m_\phi^2>\pi$.

\subsection{Finite-volume effects} 
The bound-state results discussed above are for infinite volume, we now consider how they are modified at finite volume. The bound-state solution \eq{eq:phys-solution} is exponentially localized over a scale $L_B = \kappa_B^{-1}$, so that for $L\gg L_B$, finite-volume effects are exponentially suppressed.
In the opposite regime $L\lesssim L_B$, however, the bound state does not fit entirely in the finite-volume box, and the infinite-volume solution is deformed.

\begin{figure}[h!]
\includegraphics[width=\textwidth]{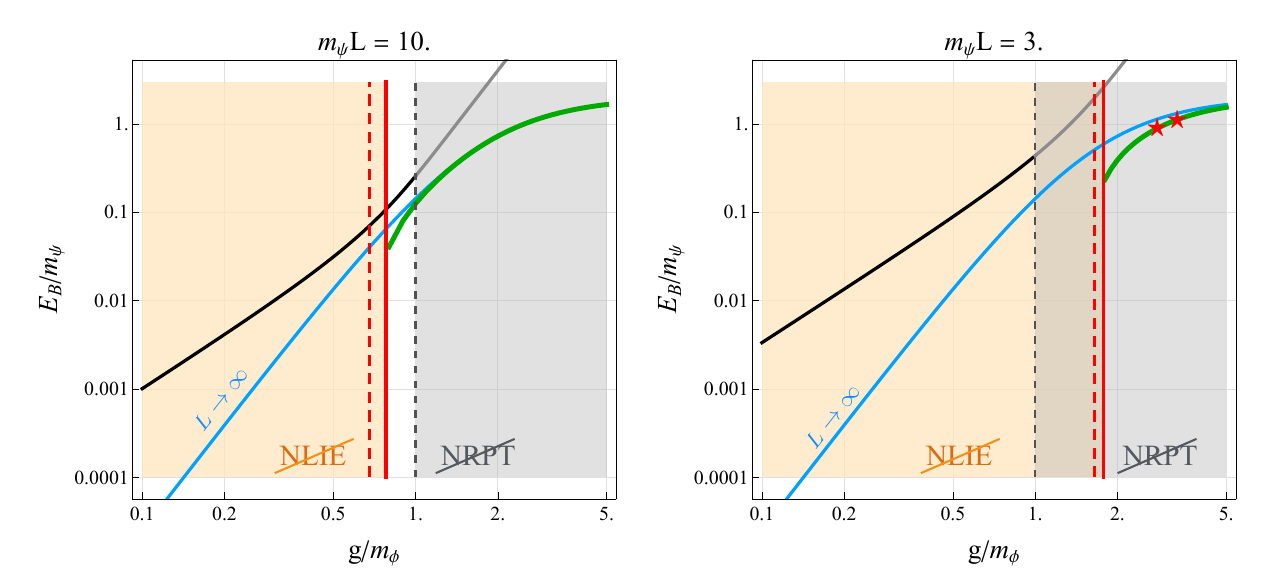}
        \caption{Finite-volume binding energy in units of $m_\psi$ in the Thirring/Sine-Gordon model computed with NRPT (black -- turning gray outside its regime of validity) and the NLIE (green), compared with the infinite-volume result (blue).
        In the gray (orange) area NRPT (the NLIE) breaks down; red solid (dashed) lines denote numerical (analytic) breakdown of the NLIE. Red stars compare our results with those of Ref.~\cite{Feverati:2000xa}. For NRPT and the Thirring/Sine-Gordon model there is no distinction between input and physical masses: $m_{\psi}=m_{\psi, \rm phys}=M_s$.}
    \label{fig:SGNR}
\end{figure}

Deep in this finite-volume regime with $L \ll L_B$, \emph{at weak coupling}, where the bound state is non-relativistic, the Yukawa potential can be treated as a perturbation of the free (discrete) finite-volume states
$\Phi_n(x)=L^{-1/2}e^{ik_n x}$ with $k_n=2\pi n/L$.
For  $m_\phi,\,g\to \infty$, with small~$\lambda \equiv g^2/m^2_\phi$ held fixed, the computations are simplified: the Yukawa potential asymptotes to a contact $\delta$-function and the box size does not affect the potential.
. 
In this limit, its expectation value between free states becomes state independent: $ V_{pq}=\matrixel{p}{V}{q}=-\lambda/L$.
Then, all dynamics are encoded in the resolvent
\begin{equation}
 S(E)\equiv \sum_{n\neq0}\frac{1}{E-E_n^{(0)}}\,, 
\end{equation}
where $E_n^{(0)}={k_n^2}/{m_\psi}$ is the non-relativistic relative energy of the fermion-antifermion system with reduced mass $m_\psi/2$. 
$S(E)$ can be computed exactly and takes qualitatively different forms depending on whether it is evaluated close to $E=0$ or not,
\begin{equation}
 S(E)=\left\{\begin{array}{ll} -\,m_\psi\left[\frac{L}{2\kappa}\coth\!\Big(\frac{\kappa L}{2}\Big)-\frac{1}{\kappa^2}\right] &\quad E=-E_B=-\kappa^2/m_\psi\\
  -\frac{m_\psi L^2}{12} & \quad E=E^{(0)}=0\,.
 \end{array}\right.
\end{equation}
With this, the entire perturbative series can be resummed, giving
\begin{equation}
 {\ E=-\frac{\lambda}{L}\frac{1}{1+\frac{\lambda}{L}S(E)}\ }.
 \label{eq:energy_rank1}
\end{equation}
  For the bound state, $E=-E_B$, and \eq{eq:energy_rank1} can be solved self-consistently, albeit numerically, as we illustrate in Fig.~\ref{fig:SGNR}.\footnote{$S(0)$ has a stronger IR divergence $\sim L^2$ than  $S(E)\sim L$ evaluated at finite $E$. In particular the combination $(\lambda/L)S(E)$ is $L$-divergent for $E=0$ but finite for $E\neq 0$. In a naive expansion around $E=0$,
  the terms are organized in such a way that at every order (from $E^{(3)}$ up) they include divergences. On the other hand, the self-consistent solution of \eq{eq:energy_rank1} has resummed these divergences to all orders.}

   Expanding \eq{eq:energy_rank1} and solving it analytically at small/large values of $L$, we find
\begin{equation}
E_B=
\left\{
\begin{array}{ll}
\displaystyle
\frac{g^2}{L m_\phi^2}
+\frac{g^4 m_\psi}{12 m_\phi^4}
+\frac{g^6 L m_\psi^2}{180 m_\phi^6}
+O\left(\frac{g^8 L^2 m_\psi^3}{m_\phi^8}\right)
&
\textrm{for}\quad L\ll \kappa_B^{-1},
\\[4mm]
\displaystyle
\frac{g^4 m_\psi}{4m_\phi^4}
+\frac{g^4 m_\psi}{m_\phi^4}
e^{-m_\psi L g^2/2m_\phi^2}
+O\left(
e^{-{m_\psi L g^2}/{m_\phi^2}}
\right)
&
\textrm{for}\quad L\gg \kappa_B^{-1}.
\end{array}
\right.
\label{eq:anapp}
\end{equation}
At the lowest orders, the $g$ and $L$ expansions overlap, so that the (necessary) $L$-resummation also affects the organization of the fixed order $g$ terms.
The small-$L$ expansion does not reproduce the coefficient of the $L$-independent $\sim g^4$ term, because it is not adapted to describe the large-$L$ regime.
Instead the large-$L$ solution converges exponentially fast in $L$ to the infinite-volume solution.  

At small $L$ the term $\sim g^2$ reappears and obscures the $\sim g^4$ infinite-volume prediction.
Moreover, the exponential large-$L$ convergence is controlled by $\kappa_B$ itself, which -- if we are to study the $g$-dependence over a large range of $g$ -- can become small and hide the clean $\sim g^4$ behavior predicted above. This will make it hard to single out the $\sim g^4$ behavior from finite-volume data.

We show the non-relativistic perturbation theory (NRPT) prediction as a black line in Fig.~\ref{fig:SGNR}, turning gray when the condition $g/m_\phi\lesssim 1$ ceases to be valid; for $m_\psi L \leq 10$, the asymptotic $\sim g^4$ behavior  sits always outside the non-relativistic approximation.

At \emph{strong coupling}, instead, the interaction cannot be treated as a perturbation of the free finite-volume states anymore. In the infinite-$m_\phi$ limit, however,  the finite-volume spectrum of the massive Thirring model can in principle  be determined through integrability of Sine-Gordon. Each energy level obeys a nonlinear integral equation (NLIE, similar in spirit to a thermodynamic Bethe ansatz) that follows from imposing periodic boundary conditions on the wavefunction~\cite{Destri:1992qk,Destri:1997yz,Feverati:1998dt}.
Carrying a particle once around the circle multiplies its wavefunction by the free propagation phase $e^{ipL}$ and by one S-matrix phase for each particle it crosses (both real and virtual). Integrability makes the scattering elastic and  the S-matrix factorizes in terms of the known infinite-volume S-matrix~\cite{Zamolodchikov:1978xm}; then the wavefunction boundary conditions  quantize the allowed energies and momenta, determining the finite-volume spectrum.
Different states are selected by explicit source terms in the quantization equation, 
together with their associated discrete quantum numbers.
The finite-volume breather masses can then be determined  numerically -- with no expansion in the coupling or in the volume -- from solutions to these equations by subtracting the finite-volume vacuum energy from the energy of a given breather state. The binding energy is then calculated using
$E_B(L)=2M_s-M_1(L)$. We  show the results in  \figref{fig:SGNR}, and provide more details on this procedure  in Appendix~\ref{app:NLIE}.

Unfortunately, the NLIE approach allows us to follow a \emph{specific state}, anchored in its $L\to \infty$ definition, rather than a specific level in the spectrum. The lowest-lying  states  that we are interested in  have  different state configurations at infinite/finite $L$. At small $L$, the level is better thought of as a fermion--antifermion pair spread over the whole volume. 
This explains why the NLIE and NRPT curves in Fig.~\ref{fig:SGNR} are not continuous: they track different states, with the NRPT state anchored at $g\to0$ and the NLIE state at $L\to\infty$, and only in the $L\to\infty$ limit do they smoothly connect. Moreover, as can be seen in the figure, at small coupling   the NLIE approach breaks down
because the simple ansatz (source term) that we use here no longer appropriately describes the level~\cite{Destri:1997yz,Feverati:1998dt,Feverati:2000xa}. 
As discussed in Appendix~\ref{app:NLIE}, the breakdown takes place at $g\lesssim g_c$, with $g_c/m_\phi\simeq 0.7(1.7)$ for $m_\psi L=10(3)$; see the red  line in \figref{fig:SGNR}.

\begin{figure}[h!]
    \centering
\includegraphics[width=0.5\linewidth]{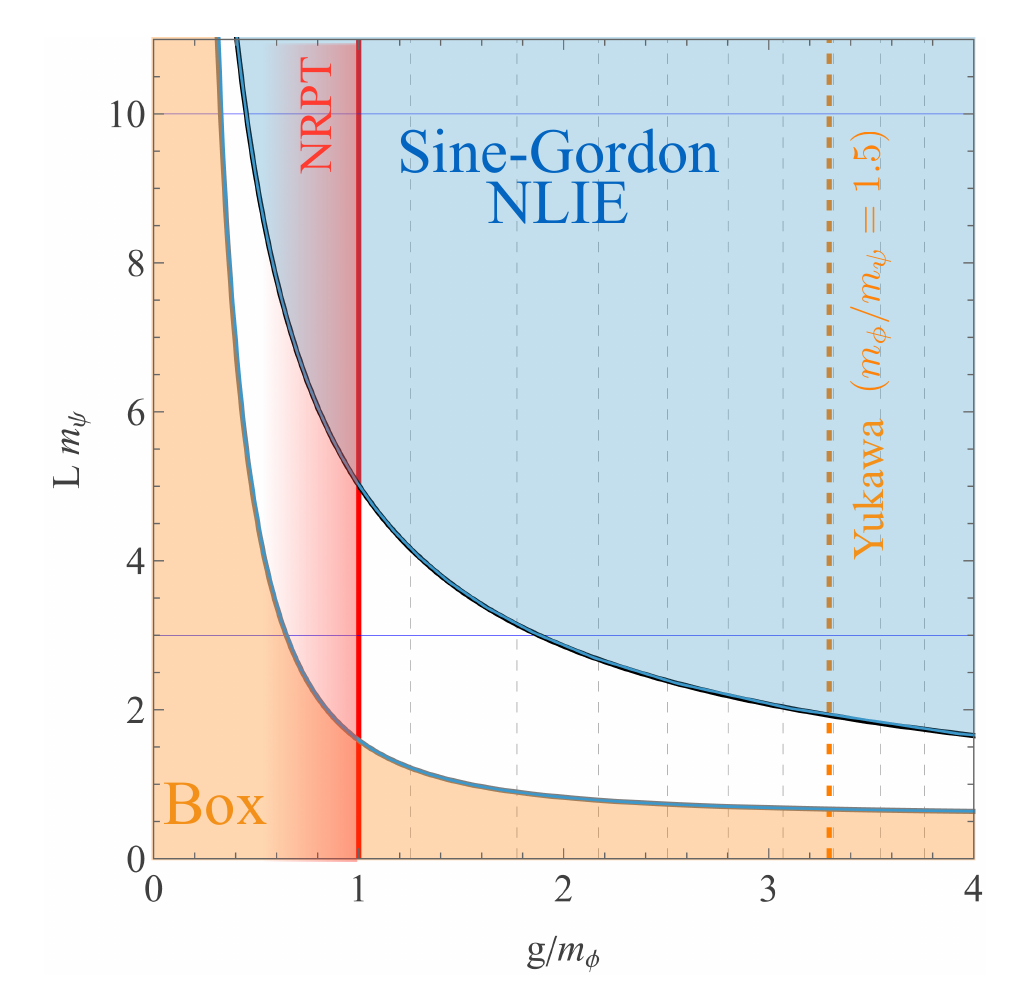}
    \caption{Sketch of the parameter-space coupling/volume, where the different approximations to compute the binding energy are reliable. At small coupling/volume $L\lesssim L_B$, the bound state doesn't fit in the box and the best approach is perturbation theory around free finite-volume states (orange area). Below $g/m_\phi \lesssim 1$ NRPT is reliable.  The NLIE results of \eq{eq:NLIE} are accurate at larger coupling/volume (in the Sine-Gordon limit),  but it breaks down at small coupling/volume, see \eq{eq:gcmono}. At large coupling the bound state might become smaller than the Yukawa reach for small enough $m_\phi$, as given by the example dashed-orange line. Light-gray vertical dashed lines correspond to thresholds where higher bound states in Sine-Gordon theory appear. }
    \label{fig:validity}
\end{figure}

In \figref{fig:validity}, we show the different regions of the volume-coupling parameter space where the different approximations hold.
Importantly, at finite volume, even in the $m_\phi\to \infty$ limit, 
neither of these methods can
compute reliably across all couplings.

\newpage
\section{Hamiltonian truncation and EFT corrections}
\label{sec:HT}
To study two-dimensional Yukawa theory across these different regimes, we use Hamiltonian truncation and compute its spectrum nonperturbatively. Splitting the Hamiltonian into a solvable part $H_0$ and an interaction term $V$, we can write it in the basis of $H_0$ eigenstates
\begin{align}
\langle f|H|i\rangle = \delta _{fi} E_i  + \langle f|V|i\rangle,
\end{align}
where $|i\rangle$ and $|f\rangle$ are eigenstates of $H_0$ with energies $E_i$ and $E_f$. The matrix $\langle f|H|i\rangle$ is infinite dimensional, but by restricting to a finite subset of states it can be diagonalized numerically to give approximate eigenvalues and eigenstates for the full theory. Hamiltonian truncation works best when the low-energy eigenstates of the full Hamiltonian $H$ are well represented by the truncated set of $H_0$ eigenstates. As the size of the truncated Hilbert space is increased, we obtain a better approximation to the full theory and a more accurate description of its low-energy physics. 

We truncate the Hilbert space by keeping states with total energy below an energy cutoff $E_{\rm max}$. This type of cutoff is intrinsically \emph{nonlocal}, as $E_{\rm max}$ is the total energy of a given state, so whether a given excitation is kept in the truncated basis depends on the energies of all the other excitations in that state. As we will see, these nonlocal effects are suppressed when $E_{\rm max}$ is large compared to the energies of the low-lying states $E_i, E_f$ and can be dealt with systematically.

For Yukawa theory we choose $H_0$ to contain the massive free scalar and massive free fermion parts of the Lagrangian \eq{eqn:Lyuk}, and $V$ to contain the interaction. We will also normal-order the Hamiltonian with respect to the massive free theory $H_0$.\footnote{To express the Lagrangian in terms of normal-ordered operators, we use
\begin{align}
\begin{split}
 \phi^2 =\ :\!\phi^2\!: + Z_\phi,\ \qquad{}
 \bar{\psi} \psi =\ :\!\bar{\psi} \psi\!: + Z_\psi
\end{split}
\end{align}
with $Z_\phi = \frac{1}{4\pi R}\sum_k \frac{1}{\omega_{\phi,k}}$ and $Z_\psi = -\frac{m_\psi}{2\pi R} \sum_{k} \frac{1 }{ \omega_k}$. The Lagrangian written in terms of normal-ordered operators is equivalent to the original Lagrangian \eq{eqn:Lyuk} up to a redefinition of the fermion mass
\begin{align}
m_\psi \rightarrow m_\psi - \frac{g^2}{m_\phi^2} Z_\psi.
\end{align}
combined with a constant shift in $\phi$. We absorb this redefinition into the definition of $m_\psi$ below.}
First, quantizing the massive scalar on a spatial circle of radius $R = L/2\pi$ gives the mode expansion,
\begin{align}
\phi(t, x) &=\frac{1}{\sqrt{2\pi R}}  \sum_k \left(\phi_k^{(-)} e^{i k_\phi \cdot x } + \phi_k^{(+)} e^{-i k_\phi \cdot x}\right),
\end{align}
with $k_\phi^\mu=\left(\omega_{\phi,k},\, k/R\right),\ x^\mu = (t, x)$ and $\eta_{\mu\nu} =\diag(1,-1)$.
The positive/negative frequency components of the field are
\begin{align}\label{eq:posneg}
\phi_k^{(-)} = \frac{1}{\sqrt{2\omega_{\phi,k}}} a_k^\dagger\ \ \textnormal{and}\ \ \phi_k^{(+)} = \frac{1}{\sqrt{2\omega_{\phi,k}}} a_k. 
\end{align}
Here $\omega_{\phi,k} = \sqrt{m_\phi^2 + k^2/R^2}$ and we have the usual commutation relation for the scalar creation and annihilation operators $[a_k,\, a_{k'}^\dagger] = \delta_{kk'}$. We also impose periodic boundary conditions
\begin{align}
\phi(t, x) =&\ \phi(t,x+2\pi R),
\end{align}
which restricts the momentum modes to $k \in \mathbb{Z}$.

Similarly, for the massive free fermion we choose periodic boundary conditions and obtain the mode expansions,
\begin{align}
\psi(t,x) &= \frac{1}{\sqrt{2\pi R}}\sum_{k}\left( \psi_k^{(-)} e^{ik \cdot x }+\psi_k^{(+)}   e^{- ik \cdot x}\right)\\
\bar{\psi}(t,x) &= \frac{1}{\sqrt{2\pi R}}\sum_{k}\left(  \bar{\psi}_k^{(-)}   e^{ i k\cdot x}+\bar{\psi}_k^{(+)}  e^{ -i k \cdot x}\right)
\end{align}
now with $k^\mu=\left(\omega_k,\, k/R\right)$ for the fermions. The positive and negative frequency components of the fields are
\begin{align}
\begin{split}
\psi_k^{(-)} &= \frac{1}{\sqrt{2\omega_{k}}} d_k^\dagger  \bm{v}_k,\qquad{} 
\psi_k^{(+)} = \frac{1}{\sqrt{2\omega_{k}}} c_k  \bm{u}_k\\ 
\bar{\psi}_k^{(-)} &= \frac{1}{\sqrt{2\omega_{k}}} c_k^\dagger \bar{\bm{u}}_k , \qquad{} 
\bar{\psi}_k^{(+)} = \frac{1}{\sqrt{2\omega_{k}}} d_k \bar{\bm{v}}_k.
\end{split}
\label{eq:fermioncomps}
\end{align}
Here $\omega_k=\sqrt{m_\psi^2+k^2/R^2}$
and the corresponding Dirac spinors are
\begin{align}
    \bm{u}_k =
    \begin{pmatrix}
        \sqrt{\omega_k-k/R}\\[4pt]
        \sqrt{\omega_k+k/R}
    \end{pmatrix}\quad{} \textnormal{and} \quad{}
    \bm{v}_k =
    \begin{pmatrix}
        \sqrt{\omega_k-k/R}\\[4pt]
        -\sqrt{\omega_k+k/R}
    \end{pmatrix}.
\end{align}
These spinors obey the standard
on-shell relations 
\begin{align}
    \left(\slashed{k}-m_\psi\right)\bm{u}_k = 0 \quad{}\textnormal{and} \quad{}
\left(\slashed{k}+m_\psi\right)\bm{v}_k &= 0
\end{align}
where $\bar{\bm{u}}_k=\bm{u}_k^\dagger\gamma^0$ and
$\bar{\bm{v}}_k=\bm{v}_k^\dagger\gamma^0$.
They are normalized so that $\bar{\bm{u}}_k \bm{u}_k  = 2m_\psi,\ \bar{\bm{v}}_k \bm{v}_k= -2m_\psi$
and satisfy the completeness relations
 $\bm{u}_k\bar{\bm{u}}_k  = \slashed{k}+m_\psi,\ \bm{v}_k\bar{\bm{v}}_k =\slashed{k}-m_\psi$. The fermion and antifermion creation and annihilation operators also obey the usual anticommutation relations $\{c_k,\, c_{k'}^\dagger\} = \delta_{kk'}$, $\{d_k,\, d_{k'}^\dagger\} = \delta_{kk'}$.  There is no spin quantum number in two dimensions so the creation/annihilation operators only have a single index corresponding to  discrete momentum. 

The normal-ordered Hamiltonian $H_0$ for the free massive scalar and free massive fermion can then be written in terms of these creation and annihilation operators as
\begin{align}
H_0 = \sum_k \left( \omega_{\phi,k}\, a_k^\dagger  a_k + \omega_k\, c_{k}^\dagger c_{k}  +\omega_k\, d^\dagger_{k} d_{k}\right).
\end{align}
The momentum and fermion charge operators are
\begin{align}
\hat{P}  = \frac{1}{R} \sum_k k \left(a^\dagger_k a_k + c_{k}^\dagger c_{k} + d^\dagger_{k} d_{k}\right),\qquad
\hat{Q} = \int dx\,\! :\!\bar{\psi} \gamma^0 \psi\!:\ =  \sum_k \left(c_{k}^\dagger c_{k} - d^\dagger_{k} d_{k}\right),
\end{align}
both of which commute with $H_0$. For the calculations in this paper we will mainly focus on states with total momentum zero, as well as the fermion charge-0 and charge-1 sectors.

A convenient choice of states is the Fock basis,
\begin{align}\label{eq:focku}
|\{n\}\rangle = |\{n_\phi, n_{\psi}, n_{\bar{\psi}}\}\rangle =&  \left( \prod_{\ell}  \frac{1}{\sqrt{n_\ell!}}\left( a_\ell^\dagger\right)^{n_\ell} c_{m_1}^\dagger\cdots c_{m_p}^\dagger  d_{k_1}^{\dagger}\cdots d_{k_q}^{\dagger}   \right) |0\rangle\, .
\end{align}
Here $n_\ell$ is the number of scalar particles with momentum $\ell$, and the fermion operators are ordered so that $m_1 < \cdots < m_p$ and $k_ 1< \cdots < k_q$.

We can now write the Yukawa interaction in terms of the same creation and annihilation operators, which act directly on the Fock basis. The normal-ordered Yukawa interaction is, 
\begin{align}
\begin{split}\label{eq:yuk}
V =&\ g \int\, dx :\!\phi \bar{\psi} \psi\!:\\
=&\ \frac{g}{4\sqrt{\pi R}} \sum_{k,\ell} \frac{1}{\sqrt{\omega_\ell\omega_k}} \Bigg[\frac{f_+(k,\ell)}{\sqrt{\omega_{\phi, \ell-k}}} \left(a_{\ell-k}^\dagger+a_{k-\ell}\right) \left(  c_{k}^\dagger  c_{\ell} +d^\dagger_{k} d_{\ell}  \right) \\
&\ \qquad{} \qquad{} \qquad{} \quad{} +\frac{f_-(k,\ell)}{\sqrt{\omega_{\phi, k+\ell}}}\left( a_{-k-\ell}^\dagger +a_{k+\ell}\right)\left( c_{k}^\dagger d^\dagger_{\ell} +d_{-k} c_{-\ell} \right)   \Bigg]
\end{split}
\end{align}
where $f_+(k,\ell) = \bar{\bm{u}}_k \bm{u}_\ell$ and $ f_-(k,\ell) = \bar{\bm{u}}_k \bm{v}_\ell$.

For the effective Hamiltonian and renormalization of the physical fermion mass, we will also need the $\phi$, $\phi^2$ and $\bar{\psi} \psi$ operators, which we write here explicitly for convenience:
\begin{align}
 \int dx\, \phi 
 =&\ \sqrt{\frac{\pi R}{ m_\phi }}\left(a_0^\dagger  + a_0\right) \\
 \int dx\, :\!\phi^2\!:\ 
 =&\  \sum_{k} \frac{1 }{2 \omega_{\phi,k}} \left[a_k^\dagger  a_{-k}^\dagger +2 a_{k}^\dagger a_k + a_k a_{-k} \right] \\
 \int dx\,:\!\bar{\psi} \psi\!:\
 =&\  \sum_{k} \frac{1 }{ \omega_k}\left[ m_\psi \left( c^\dagger_{k}   c_{k}    +d_{k}^\dagger d_{k}   \right) +\frac{k}{R}\left( c_{-k}d_{k}    + c^\dagger_{k} d_{-k}^\dagger  \right) \right].
\end{align}

\subsection{Effective Hamiltonian and power counting}\label{sec:EFT}
Truncating the spectrum to $E\leq E_{\rm max}$  introduces errors that can be systematically reduced using an EFT approach \cite{Cohen:2021erm}.
We assume that the effective Hamiltonian $H_{\rm eff}$, which acts only on the truncated set of states, can be written as
\begin{align}
H_{\rm eff} = \sum_n H_n
\label{eq:Heff}
\end{align}
with $H_1=V$; for $n\geq2$, $H_n\sim O(V^n)$ contains both local and nonlocal interactions. The nonlocality comes from terms in the $E/E_{\rm max}$ expansion: when measured with respect to $H_0$ these terms  appear as powers of $H_0$ multiplying local operators, schematically $\int dx\, H_0^p\, {\cal O}$.
The decomposition \eq{eq:Heff} is particularly useful when all low-energy scales are much smaller than $E_{\rm max}$, so that only a finite number of terms need to be kept at a given order in $1/E_{\rm max}$.
Following~\cite{Cohen:2021erm}, we determine the terms $H_n$ by matching the overlap of the free and interacting low-energy states in the full and truncated theories order by order. Schematically, we can identify and power-count the following terms at each order in $g$:
\begin{align}
H_{2} &\sim g^2  \int dx \bigg[ \mathbbm{1}\ln(E_{\rm max}) +\phi^2 \ln(E_{\rm max})+ \frac{H_0}{E_{\rm max}} + \frac{H_0}{E_{\rm max}} \phi^2+\frac{m_\phi^2}{E^2_{\rm max}} \phi^2+ \frac{m_\psi}{E_{\rm max}^2} \bar{\psi} \psi  + \cdots\bigg]\nonumber\\ \label{eq:Heffsch}
H_{3} &\sim \frac{g^3}{E_{\rm max}}  \int dx \bigg[\frac{1}{E_{\rm max}}\phi \bar{\psi} \psi  + \frac{m_\psi}{E_{\rm max}}\phi^3+ \frac{1}{E_{\rm max}}H_0\phi^2 +\cdots\bigg]\\
H_{4} &\sim \frac{g^4}{E_{\rm max}^2}  \int dx \bigg[\mathbbm{1} + \phi^2 + \frac{m_\psi}{E_{\rm max}^2} \bar{\psi} \psi  + \frac{1}{E_{\rm max}}H_0\phi^2 + \cdots\bigg],\nonumber
\end{align}
where each term is multiplied by a different coefficient, which we do not show for compactness. These coefficients may also depend logarithmically on $E_{\rm max}$; we have shown this explicitly only for the first two terms to highlight their divergence.
The allowed terms and their scaling follow from naive dimensional analysis and the symmetries of the theory (e.g.~$\bar\psi\psi$ must appear with a power of $m_\psi$). Immediately we see that there are two possible UV divergences: one for the vacuum energy and one for the scalar mass. As these are logarithmic divergences (rather than power-law), they can be compensated for using local counterterms in equal-time Hamiltonian truncation \cite{Rutter:2018aog}.
To cancel these divergences, we follow the procedure of \cite{Cohen:2021erm, Delouche:2023wsl}: first adding local counterterms to the full theory, and then using the matching calculation at the appropriate order ($g^2$ in this case) to determine the corresponding effective Hamiltonian terms.

This power counting tells us that, before adding counterterms and effective Hamiltonian corrections, some observables may depend logarithmically on the cutoff $E_{\rm max}$. If we then correctly add the counterterms and corrections corresponding to the first four terms in $H_2$, the leading remaining cutoff dependence should scale as $1/E_{\rm max}^2$.

\subsection{Matching}
\label{sec:match}

To determine the effective  Hamiltonian, we match the overlap between the free and interacting low-energy eigenstates in the full and truncated theories. The interacting eigenstates are precisely the states we want the truncated theory to reproduce, and matching their overlaps with the $H_0$ eigenstates ensures that the low-energy eigenstates are well represented in the truncated Hilbert space, order by order in $V$.

To relate an $H_0$ eigenstate to an interacting eigenstate, we use a construction inspired by Gell-Mann-Low to adiabatically connect the free and
interacting theories
\cite{Gell-Mann:1951ooy,Molinari_2007}. We use the time-dependent Hamiltonian
\begin{equation}
H_\epsilon = H_0 + e^{-\epsilon |t|}V ,
\end{equation}
which approaches $H_0$ as $t\to\pm\infty$ and the full interacting Hamiltonian
at $t=0$.

The corresponding interaction-picture time evolution operator is
\begin{align}
U_{\rm IP, \epsilon}(t_f,t_i) = T\exp\left[-i \int_{t_i}^{t_f}\, dt\ e^{i H_0 t} e^{-\epsilon |t|} V e^{-i H_0t} \right].
\end{align}
For the matching calculation we need the overlap between an interacting eigenstate $|\Psi_f\rangle$ and an $H_0$ eigenstate
$|i\rangle$. At finite
$\epsilon$, we calculate this overlap as
\begin{align}
\langle\Psi_{f,\epsilon}|i\rangle
=
\lim_{t_f\rightarrow\infty}
\langle f|U_{\rm IP,\epsilon}(t_f,0)|i\rangle
\end{align}
 which we can then match  between the full and truncated theories:
\begin{align}
\langle \Psi_{f,\epsilon} |i \rangle_{\textnormal{full}} =\langle \Psi_{f,\epsilon} |i \rangle_{\textnormal{eff}}. 
\end{align}
Here $\epsilon$ is still finite and acts as an IR regulator, which we send to zero after the matching calculation is performed. 
In this limit, $|\Psi_{f,\epsilon}\rangle$ becomes the interacting scattering state
associated with the free asymptotic state $|f\rangle$. The matching we do is therefore closely related to the S-matrix matching used in standard EFT. Computing this overlap for the \emph{full theory} order by order in $V$ gives
\begin{align}
\langle \Psi_{f,\epsilon} |i \rangle_{\textnormal{full}} = \delta_{fi} + \frac{\langle f|V|i\rangle}{E_{fi} + i\epsilon} + \sum_\alpha \frac{\langle f |V|\alpha\rangle \langle \alpha |V|i \rangle}{(E_{fi} + i\epsilon)(E_{f\alpha} + i \epsilon)} + \mathcal{O}(V^3)
\label{eq:matchfull}
\end{align}
where $E_{fi} \equiv E_f - E_i$ and $|i\rangle$ ($|f\rangle$) are the ``initial'' (final) eigenstates of $H_0$. Performing the same calculation for the effective Hamiltonian, which has the form \eq{eq:Heff}, gives,
\begin{align}
\langle \Psi_{f,\epsilon} |i \rangle_{\textnormal{eff}} = \delta_{fi} + \frac{\langle f|H_1|i\rangle}{E_{fi} + i\epsilon}+ \frac{\langle f|H_2|i\rangle}{E_{fi} + i\epsilon} + \sum^<_\alpha \frac{\langle f |H_1|\alpha\rangle \langle \alpha |H_1|i \rangle}{(E_{fi} + i\epsilon)(E_{f\alpha} + i \epsilon)} + \mathcal{O}(V^3). 
\label{eq:matcheff}
\end{align}
The effective Hamiltonian only acts on the truncated set of states with $E_\alpha \leq E_{\rm max}$, and we denote sums over these states by $\sum^<_\alpha$. Equating \eq{eq:matchfull} and \eq{eq:matcheff} at each order in the interaction gives us a unique expression for matrix elements of our effective Hamiltonian:
\begin{align}
\langle f|H_1|i\rangle &= \langle f|V|i\rangle\\
\langle f|H_2|i\rangle &= \sum^>_\alpha \frac{\langle f |V|\alpha\rangle \langle \alpha |V|i \rangle}{E_{f\alpha}}\label{eq:H2}\\
\vdots\nonumber
\end{align}
where $\sum^>_\alpha$ indicates summing over states with energy $E_\alpha > E_{\rm max}$ and we have safely taken $\epsilon \rightarrow 0$ at the end of the matching calculation.  This choice of matching condition gives a systematic expansion in which separation of scales is maintained even at higher orders \cite{Cohen:2021erm,Demiray:2025zqh}. The intermediate states in Eq.~\eqref{eq:H2} have energies $E_\alpha>E_{\rm max}$, while the external energies and other IR scales are taken to be much smaller than $E_{\rm max}$, enabling the $1/E_{\rm max}$ expansion used below to compute the effective Hamiltonian. One consequence of this choice is that the effective Hamiltonian is non-Hermitian at finite $E_{\rm max}$, since the denominator depends on  $E_f$ rather than symmetrically on $E_i$ and~$E_f$. In practice, the low-lying eigenvalues we compute are real to numerical precision.

\subsubsection{UV divergences and counterterms}
\label{sec:CT}

So far the procedure has closely followed \cite{Cohen:2021erm} in the construction of the effective Hamiltonian. However, unlike the previous case, Yukawa theory in two dimensions has UV divergences that must be accounted for with counterterms before truncation. Because the~$E_{\rm max}$ cutoff is necessarily nonlocal, sufficiently severe UV divergences generally require counterterms with nonlocal components in order to recover local physics as $E_{\rm max} \rightarrow\infty$~\cite{Rutter:2018aog}. In this case, however, the divergences are all logarithmic in the cutoff, so equal-time Hamiltonian truncation requires only local counterterms.\footnote{This differs from lightcone quantization, where nonlocal, state-dependent counterterms can be required even for logarithmic UV divergences \cite{Anand:2020qnp}.} Building on HTET, recent work developed a closely related matching procedure for theories with nontrivial UV divergences \cite{EliasMiro:2022pua,Delouche:2023wsl,Delouche:2024yuo}. In this approach, the full theory is first renormalized using a local regulator before matching onto the truncated Hilbert space. This procedure has been successfully tested in bosonic theories defined as relevant deformations of a CFT, where it reproduces the expected RG flows between minimal models. Here we adopt the same prescription of renormalizing the theory before matching, but compute the counterterms directly within the diagrammatic framework of HTET, extending the method to a massive theory with both fermionic and bosonic degrees of freedom and nontrivial UV divergences.

In this theory there are two UV divergences, one in the vacuum energy and one in the scalar self-energy, both at $\mathcal{O}(g^2)$. These divergent contributions can be calculated in old-fashioned (Hamiltonian) perturbation theory either by brute force or using the diagrammatic rules presented in Appendix \ref{app:Feyn}. We first regulate the theory using a high-energy Wilsonian cutoff $\Lambda> E_{\rm max}$, and we will see that the effective Hamiltonian computed correctly will have a finite $\Lambda \rightarrow \infty$ limit.

First, using \eq{eq:yuk}, we calculate the contribution to the vacuum energy in the full theory,
\begin{align}
\begin{split}
\includegraphics[scale=0.45,trim={0cm 0cm 0cm 0cm},clip,valign=c]{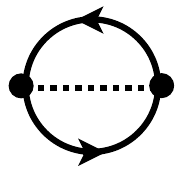} 
=&\ \frac{-g^2}{8\pi R} \langle f |i\rangle \sum^{\Lambda R}_{k,\ell = -\Lambda R} \frac{1}{\omega_k\omega_\ell\omega_{\phi,k+\ell}} \frac{\left(\omega_k \omega_\ell-m^2_\psi-\frac{k\ell}{R^2}\right)}{ (\omega_k +\omega_\ell +\omega_{\phi, k+\ell}- i\epsilon)}.
\label{eq:vac}
\end{split}
\end{align}
The sum here is independent of the external states $|i\rangle$ and $|f\rangle$, and we can convert this expression into a coefficient for a local operator in the Hamiltonian,  $\mathbbm{1}$, using $\langle f| \int dx\, \mathbbm{1} |i \rangle = 2\pi R\langle f|i \rangle$. Once we have truncated the theory to finite $E_{\rm max}$, these sums can depend nontrivially on the external state energy $E_f$ or $E_i$, giving nonlocal contributions to our effective Hamiltonian. These nonlocal terms are generically suppressed with $E_{\rm max}$ \cite{Rutter:2018aog}, but are necessary at higher order in the effective Hamiltonian to compensate for the nonlocal nature of the energy cutoff $E_{\rm max}$.

Performing a similar calculation for the scalar self-energy, we get
\begin{align}
\begin{split}
\includegraphics[scale=0.45,trim={0.5cm 0cm 0.5cm 0cm},clip,valign=c]{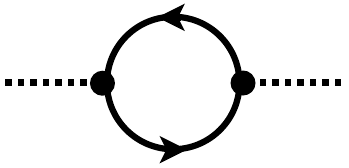}+\includegraphics[scale=0.45,trim={0cm 0cm 0cm 0cm},clip,valign=c]{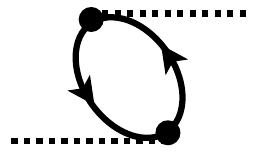} 
=&\ \frac{-g^2}{2\pi R} \langle f |\phi_{0}^{(-)} \phi_{0}^{(+)}|i\rangle\sum^{\Lambda R}_{k = -\Lambda R} \frac{1}{\omega_k}\frac{4k^2/R^2}{4\omega^2_k - m^2_\phi-4i\epsilon \omega_k}.
\end{split}\label{eq:phi2}
\end{align}
We have set the external momentum to zero, and the notation $\phi_0^{(\pm)}$ refers to the positive/negative frequency components of the field $\phi$ defined in \eq{eq:posneg}. Once again, the factor $\langle f|\phi^{(-)}_0\phi^{(+)}_0|i\rangle$ can be converted into the matrix element of the local operator $\phi^2$ using $\langle f|\int dx\,:\!\phi^2\!:|i\rangle =  2\langle f|\phi^{(-)}_0\phi^{(+)}_0|i\rangle$ for this choice of external states. In this case there are two diagrams that differ by the vertex ordering/intermediate state composition. While each individual diagram's contribution to the scalar self-energy is not covariant, the sum of them is, as expected in old-fashioned perturbation theory. This is no longer the case once the Hamiltonian is truncated to states with finite $E_{\rm max}$, and will lead to the dependence on the external state energy and nonlocal effects at subleading order in  the $1/E_{\rm max}$ expansion.

We then add local counterterms to our Hamiltonian to cancel these divergences, so that our full Hamiltonian is now
\begin{align}
H = H_0 + V + H_{\rm CT}.
\label{eq:H+CT}
\end{align}
The counterterm Hamiltonian consists of two parts
\begin{align}
H_{\rm CT,\, \mathbbm{1}} &= \frac{g^2}{16\pi^2 R^2} \sum^{\Lambda R}_{k,\ell = -\Lambda R} \frac{1}{\omega_k\omega_\ell\omega_{\phi,k+\ell}} \frac{\left(\omega_k \omega_\ell-m^2_\psi-\frac{k\ell}{R^2}\right)}{(\omega_k + \omega_\ell +\omega_{\phi, k+\ell}-i\epsilon)}\int dx\, \mathbbm{1} \\
H_{\rm CT,\, \phi^2} &= \frac{g^2}{4\pi R}\sum^{\Lambda R}_{k = -\Lambda R}\frac{k^2/R^2}{\omega^3_k} \int dx\, :\!\phi^2\!:. 
\end{align}
For the $\phi^2$ counterterm we have chosen a renormalization scheme that cancels only the most divergent part of \eq{eq:phi2}. With the addition of these counterterms, the full theory is now UV-finite. This is the full Hamiltonian we will match to in order to calculate the effective Hamiltonian corrections. 

\subsubsection{$H_{\rm eff}$}
\label{sec:Heff}

We can now calculate our effective Hamiltonian $H_{\rm eff}$ by matching onto the finite, renormalized Hamiltonian \eq{eq:H+CT}. According to the power counting in Sec.~\ref{sec:EFT}, to improve convergence up to $1/E_{\rm max}^2$, we should include the first four terms in $H_2$ in \eq{eq:Heffsch}, which are the contributions to operators $\mathbbm{1}$ and $\phi^2$ at leading and next-to-leading order in the $1/E_{\rm max}$ expansion. 

Matching the full and truncated theories gave us the general expression for $H_2$ in \eq{eq:H2}
\begin{align}
\langle f|H_2|i\rangle = \sum_\alpha^> \frac{\langle f|V|\alpha\rangle \langle \alpha|V|i \rangle}{E_{f\alpha}}.
\end{align}
Applying this to the vacuum state gives 
\begin{align}
\langle f|H_2|i\rangle_{\mathbbm{1}} =&\ \Bigg[\includegraphics[scale=0.45,trim={0cm 0cm 0cm 0cm},clip,valign=c]{Figs/Yukawa-VacBubble}\Bigg]_{\rm full}-\Bigg[\includegraphics[scale=0.45,trim={0cm 0cm 0cm 0cm},clip,valign=c]{Figs/Yukawa-VacBubble}\Bigg]_{\rm eff}\\ 
=&\ \frac{-g^2}{8\pi R} \langle f|i\rangle\sum^{\Lambda R}_{k,\ell = -\Lambda R} \frac{\Theta(E_f +\omega_{\phi, k+\ell} +\omega_k + \omega_\ell - E_{\rm max})}{\omega_k\omega_\ell\omega_{\phi,k+\ell}} \frac{\left(\omega_k \omega_\ell-m^2_\psi-\frac{k\ell}{R^2}\right)}{ (\omega_k +\omega_\ell +\omega_{\phi, k+\ell})}\nonumber. 
\end{align}
The $\Theta$ term in this sum enforces that the intermediate state energy $E_\alpha > E_{\rm max}$, coming from the difference between the full theory sum over all states and the truncated theory sum over states with $E_\alpha \leq E_{\rm max}$. Here $E_f$ is the total final state energy, which in this vacuum case is made up entirely of spectator particles, which we have omitted from the diagrammatic pictures. 
We can then expand this expression in the ratio of low-energy scales to $E_{\rm max}$ assuming
\begin{align}\label{eq:LEexp}
\omega_{i, f} \leq E_{i,f} \ll E_{\rm max}
\end{align}
where $\omega_{i,f}$ is the energy of individual external particles and $E_{i,f}$ is the total energy of the external states. This  is  valid  only for the lowest energy states in our system, as we would expect for a low-energy effective theory.

The first two terms in this expansion for the vacuum are 
\begin{align}
\begin{split}
\langle f|H_2|i\rangle_{\mathbbm{1}} 
\approx&\ \frac{-g^2}{8\pi R} \langle f|i\rangle\sum^{\Lambda R}_{k,\ell = -\Lambda R} \frac{\Theta(\omega_{\phi, k+\ell} +\omega_k + \omega_\ell - E_{\rm max})}{\omega_k\omega_\ell\omega_{\phi,k+\ell}} \frac{\left(\omega_k \omega_\ell-m^2_\psi-\frac{k\ell}{R^2}\right)}{ (\omega_k +\omega_\ell +\omega_{\phi, k+\ell})}\\
&-E_f \frac{g^2}{4\pi E_{\rm max}}\langle f|i\rangle
\sum_{k = -\Lambda R}^{\Lambda R} \frac{\Theta( E_{\rm max}-|k|/R - \omega_{\phi,k})}{\omega_{\phi,k}}
+\mathcal{O}(1/E_{\rm max}^2)
\end{split}
\end{align}
The first term gives a contribution to the identity operator, while the second term proportional to $E_f$ can be repackaged at operator-level as a $1/E_{\rm max}$ contribution to $H_0$.

Similarly we can compute the contribution to $H_2$ from the scalar self-energy diagrams
\begin{align}
\begin{split}
\langle f|H_2|i\rangle_{\phi^2} =&\ \Bigg[\includegraphics[scale=0.45,trim={0.5cm 0cm 0.5cm 0cm},clip,valign=c]{Figs/Yukawa-ScalarMass1} + \cdots\Bigg]_{\rm full}-\bigg[\includegraphics[scale=0.45,trim={0.5cm 0cm 0.5cm 0cm},clip,valign=c]{Figs/Yukawa-ScalarMass1} + \cdots\bigg]_{\rm eff}\\
=&\ -\frac{g^2}{2\pi R} \langle f |\phi_{p_2}^{(-)} \phi_{p_1}^{(+)}|i\rangle \sum^{\Lambda R}_{k = -\Lambda R} \frac{\Theta(2\omega_k -E_{\rm max})}{\omega_k^3}\frac{k^2}{R^2}\\
&-\frac{1}{2} E_{fi}\frac{g^2}{4\pi R} \langle f |\phi_{p_2}^{(-)} \phi_{p_1}^{(+)}|i\rangle \sum^{\Lambda R}_{k = -\Lambda R} \frac{\Theta(2 \omega_k -E_{\rm max})}{\omega_k^4} \frac{k^2}{R^2}\\
&-\frac{1}{2}\left(E_i + E_f\right) \frac{g^2}{\pi E_{\rm max} } \langle f |\phi_{p_2}^{(-)} \phi_{p_1}^{(+)}|i\rangle + \mathcal{O}(1/E_{\rm max}^2).
\end{split}
\label{eq:H2scalar}
\end{align}

The first term in this expansion gives a local correction to $\phi^2$, whereas the last two terms are \emph{nonlocal}. These correspond to either $H_0 \times \int dx\, \phi^2$ ($E_f$ terms) or $\int dx\, \phi^2 \times H_0$ ($E_i$ terms). Since $H_0$ itself is already the integral of a local operator, this combination is nonlocal. These terms come entirely from $E_{i,f}$ in this expansion, and we generically expect all nonlocal terms in the effective Hamiltonian to be of the schematic form $H^n_0 \times \mathcal{O}$, 
where $\mathcal{O}$ is a local operator in the theory (possibly including derivatives). This also makes it clear that these terms must appear with powers of $1/E_{\rm max}^n$ and vanish in the limit $E_{\rm max} \rightarrow \infty$. 

At leading order in the $1/E_{\rm max}$ expansion we have:
\begin{align}
H^{(\rm LO)}_{\rm eff} = H_0 + V + H_{\rm CT}+ H^{(\rm LO)}_{2}
\end{align}
with the  local effective Hamiltonian terms calculated at order $g^2$
\begin{align}
H_{2,\mathbbm{1}}^{\rm (LO)} =&\  \frac{-g^2}{16\pi^2 R^2}\sum^{\Lambda R}_{k,\ell = -\Lambda R} \frac{\Theta( \omega_k + \omega_\ell + \omega_{\phi,k+\ell}-E_{\rm max} )}{\omega_k\omega_\ell\omega_{\phi,k+\ell}} \frac{\left(\omega_k \omega_\ell-m^2_\psi  -\frac{k\ell}{R^2}\right)}{ (\omega_k + \omega_\ell +\omega_{\phi, k+\ell})}\int dx\, \mathbbm{1}\\
H_{2,\phi^2}^{(\rm LO)} =&\  -\frac{g^2}{4\pi R} \sum^{\Lambda R}_{k = -\Lambda R} \frac{\Theta(2 \omega_k  - E_{\rm max})}{\omega_k^3}\frac{k^2}{R^2} \int dx\, :\!\phi^2\!:.
\end{align}
These combine with $H_{\rm CT}$ to give the effective Hamiltonian at leading order
\begin{align}
H^{(\rm LO)}_{\rm eff} = H_0 + V +\mathcal{C}^{\rm(LO)}_{\phi^2}  \int dx\, :\!\phi^2\!: + \ \mathcal{C}^{\rm(LO)}_{\mathbbm{1}}\int dx\, \mathbbm{1}
\end{align}
with
\begin{align}\label{eq:phi2lo}
\mathcal{C}^{\rm(LO)}_{\phi^2} 
=&\ \frac{g^2}{4\pi R} \sum_{k} \frac{\Theta(E_{\rm max}-2 \omega_k)}{\omega_k^3}\frac{k^2}{R^2}\\
\mathcal{C}^{\rm(LO)}_{\mathbbm{1}} 
=&\ \frac{g^2}{16\pi^2 R^2} \sum_{k,\ell} \frac{\Theta( E_{\rm max} -\omega_k - \omega_\ell - \omega_{\phi,k+\ell})}{\omega_k\omega_\ell\omega_{\phi,k+\ell}} \frac{\left(\omega_k \omega_\ell-m^2_\psi  -\frac{k\ell}{R^2}\right)}{ \omega_k + \omega_\ell +\omega_{\phi, k+\ell}}.
\label{eq:ident}
\end{align}
These coefficients are now finite, as they include only contributions from energies below $E_{\rm max}$ and we have safely taken the $\Lambda \rightarrow \infty$ limit. Although we kept terms proportional to the identity to illustrate the procedure, we drop them from now on since they do not contribute to energy differences. 

Adding in the $1/E_{\rm max}$ corrections from \eq{eq:H2scalar} gives the final NLO result:
\begin{align}
H^{(\rm NLO)}_{\rm eff} = H_0 + V + H_{\rm CT}+ H^{(\rm LO)}_{2}+H^{(\rm NLO)}_{2}
\end{align}
which we can write out explicitly as
\begin{align}
\begin{split}
H^{(\rm NLO)}_{\rm eff} =&\ (1+\mathcal{C}^{\rm (NLO)}_{\mathbbm{1},1}) H_0 + V +\mathcal{C}^{\rm (LO)}_{\phi^2} \int dx\, :\!\phi^2\!:\\
&+ \mathcal{C}^{\rm (NLO)}_{\phi^2,\, 1} \left\{H_0, \int dx\, :\!\phi^2\!:\right\} +  \mathcal{C}^{\rm (NLO)}_{\phi^2,\, 2} \left[H_0, \int dx\, :\!\phi^2\!:\right]
\end{split}
\label{eq:HeffNLO}
\end{align}

with the additional coefficients
\begin{align}
\label{eq:HTETcoeffs1}
\mathcal{C}^{\rm (NLO)}_{\phi^2,\, 1}=&\  -\frac{g^2}{4 \pi E_{\rm max}}\\
\label{eq:HTETcoeffs2}
\mathcal{C}^{\rm (NLO)}_{\phi^2,\, 2} =&\  -\frac{g^2}{16 \pi R} \sum_{k} \frac{\Theta(2 \omega_k -E_{\rm max})}{\omega_k^4} \frac{k^2}{R^2}\\
\label{eq:HTETcoeffs3}
\mathcal{C}^{\rm (NLO)}_{\mathbbm{1},1} =&\ - \frac{g^2}{4\pi E_{\rm max}}
\sum_{k} \frac{\Theta( E_{\rm max}-|k|/R - \omega_{\phi,k})}{\omega_{\phi,k}}. 
\end{align}
This is our complete effective Hamiltonian including terms up to $O(1/E_{\max})$, with remaining corrections of $O(1/E_{\max}^2)$.

\subsubsection{Physical renormalization conditions}\label{sec:massRG}

Although the effective theory \eq{eq:HeffNLO} is now technically finite, it is defined in terms of the UV input parameters. In practical applications it is useful to have direct control of the physical output parameters, such as the physical masses of fermion or scalar one-particle states. Akin to ordinary renormalization in a local QFT, this requires explicit mass term contributions to be treated as interactions -- in Hamiltonian truncation this will enable us to numerically and nonperturbatively fix the mass, as the other parameters in the theory are varied. Similar numerical tuning of the input masses was used in the early Hamiltonian studies of this model~\cite{Brooks:1983sb,Pauli:1985ps}. We show here what this implies for the fermion mass. Although not done in this work, in principle we could use the same procedure to fix the physical scalar mass. However, at the order in $1/E_{\rm \max}$ considered here, this would not require any new correction terms, as they would enter at $O(1/E_{\rm max}^2)$.

To fix the physical fermion mass numerically, we add the interaction term
\begin{align}
\Delta V = m_{V,\psi} \int dx\, :\!\bar{\psi} \psi\!: 
\end{align}
to the Hamiltonian. Diagrammatically, this is an additional interaction vertex of the form
\begin{equation}
\includegraphics[scale=0.5,trim={0cm 0cm 0cm 0cm},clip,valign=c]{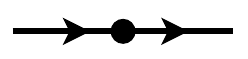} = m_{V,\psi}.
\label{eq:mvpsirule}
\end{equation}
The coupling $m_{V,\psi}$ has the same mass dimension as $g$, so power counting tells us we only need to include terms from $H_2$ in our effective Hamiltonian to have the same level of convergence as before. 
The new possible terms in $H_2$ with errors of $\mathcal{O}(1/E_{\rm max}^2)$ are:
\begin{align}
\begin{split}
\Delta H_{2} \sim &\,   g\, m_{V,\psi} \int dx \bigg[ \phi \ln (E_{\rm max})+ \frac{H_0}{E_{\rm max}} \phi + \mathcal{O}(1/E_{\max}^2)\bigg]\\
&+m_{V,\psi}^2 \int dx \bigg[ \mathbbm{1} \ln (E_{\rm max}) + \frac{H_0}{E_{\rm max}} \mathbbm{1} + \mathcal{O}(1/E_{\max}^2)  \bigg].
\end{split}
\end{align}
There are two new UV divergences associated with this interaction, one in the vacuum energy and one from the scalar tadpole. We can calculate these contributions diagrammatically using the rules of Appendix \ref{app:Feyn}. For the vacuum we find
\begin{equation}
\includegraphics[scale=0.5,trim={0cm 0cm 0cm 0cm},clip,valign=c]{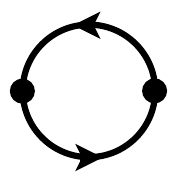} = -\frac{1}{2}m_{V,\psi}^2\langle f|i\rangle \sum_{k = -\Lambda R}^{\Lambda R} \frac{k^2/R^2}{\omega_k^3}. 
\end{equation}
For the scalar tadpole, there are two diagrams with different vertex ordering/intermediate states that contribute 
\begin{equation}
\includegraphics[scale=0.5,trim={0cm 0cm 0cm 0cm},clip,valign=c]{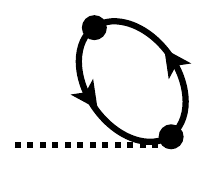}+\includegraphics[scale=0.5,trim={0cm 0cm 0cm 0cm},clip,valign=c]{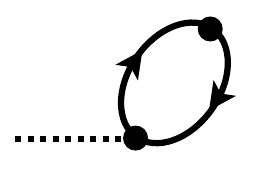}  = -\frac{g m_{V,\psi}}{\sqrt{2\pi R}} \langle f |\phi_{0}^{(-)}|i\rangle\sum_{k = -\Lambda R}^{\Lambda R} \frac{k^2/R^2}{2\omega_k^3} \frac{(4\omega_k-m_\phi)}{(2\omega_k-m_\phi )}.
\end{equation}
To cancel the divergent parts of these diagrams, we follow the same procedure as before and introduce local counterterms into the full theory, 
\begin{align}
\Delta H_{\rm CT, \mathbbm{1}} =&\ \frac{m_{V,\psi}^2}{4\pi R} \sum_{k = -\Lambda R}^{\Lambda R} \frac{k^2/R^2}{\omega^3_k}\int dx\, \mathbbm{1}\\
\Delta H_{\rm CT, \phi}=&\ \frac{g m_{V,\psi} }{2\pi R} \sum_{k = -\Lambda R}^{\Lambda R}  \frac{k^2/R^2}{\omega_k^3}\int dx\, \phi.
\end{align}
 The full Hamiltonian including counterterms is then
\begin{align}
H_{m_\psi} = H_0 + V + \Delta V + H_{\rm CT} + \Delta H_{\rm CT}.
\end{align}
We can now perform a matching calculation using this finite Hamiltonian to find our new effective Hamiltonian corrections. 

For the vacuum state, following the same prescription as the Yukawa interaction, we find 
\begin{align}
\langle f|\Delta H_2|i\rangle_{\mathbbm{1}} =&\ \Bigg[\includegraphics[scale=0.5,trim={0cm 0cm 0cm 0cm},clip,valign=c]{Figs/Yukawa-VacuumFermionVertex}\Bigg]_{\rm full}-\Bigg[\includegraphics[scale=0.5,trim={0cm 0cm 0cm 0cm},clip,valign=c]{Figs/Yukawa-VacuumFermionVertex}\Bigg]_{\rm eff}\\ 
\approx &\ -\frac{1}{2} m_{V,\psi}^2\langle f|i\rangle \sum_{k = -\Lambda R}^{\Lambda R} \frac{\Theta( 2\omega_k - E_{\rm max})}{\omega_k^3}\frac{k^2}{R^2}\\
&- m_{V,\psi}^2\frac{ R E_f}{E_{\rm max}} \langle f|i\rangle + \mathcal{O}(1/E_{\rm max}^2).\nonumber
\end{align}
Finally, for the effective Hamiltonian associated with the scalar tadpole, 
\begin{align}
\langle f|\Delta H_2|i\rangle_{\phi} =&\ \Bigg[\includegraphics[scale=0.5,trim={0cm 0cm 0cm 0cm},clip,valign=c]{Figs/Yukawa-ScalarFermionVertex2}+\cdots\Bigg]_{\rm full}-\Bigg[\includegraphics[scale=0.5,trim={0cm 0cm 0cm 0cm},clip,valign=c]{Figs/Yukawa-ScalarFermionVertex2}+\cdots\Bigg]_{\rm eff}\\ 
\approx&\ - \frac{g m_{V,\psi}}{\sqrt{2\pi R}} \langle f |\phi_{0}^{(-)}|i\rangle\sum_{k = -\Lambda R}^{\Lambda R} \frac{k^2}{R^2} \frac{\Theta(2\omega_k - E_{\rm max})}{\omega^3_k}\\
& -\frac{1}{4} E_{fi}\, \frac{g m_{V,\psi}}{\sqrt{2\pi R}} \langle f |\phi_{0}^{(-)}|i\rangle \sum_{k = -\Lambda R}^{\Lambda R} \frac{k^2}{R^2} \frac{\Theta(2\omega_k - E_{\rm max})}{\omega^4_k}\nonumber\\
&- R (E_i + E_f) \frac{g m_{V,\psi}}{\sqrt{2\pi R}} \langle f |\phi_{0}^{(-)}|i\rangle\frac{1}{E_{\rm max}} + \mathcal{O}(1/E_{\rm max}^2)\,.\nonumber
\end{align}
With these corrections, our full effective Hamiltonian including fermion mass renormalization at leading order is 
\begin{align}
H_{\rm eff,\, m_\psi}^{(\rm LO)} =  H_0 + V + \Delta V + H_{\rm CT} +\Delta H_{\rm CT} + H_2^{(\rm LO)} + \Delta H_2^{\rm (LO)}
\end{align}
with the effective Hamiltonian corrections
\begin{align}
\Delta H_{2,\mathbbm{1}}^{(\rm LO)} &= -\frac{m_{V,\psi}^2}{4\pi R}\sum_{k = -\Lambda R}^{\Lambda R}\frac{k^2}{R^2} \frac{\Theta(2\omega_k-E_{\rm max})}{\omega^3_k}\int dx\, \mathbbm{1}\\
\Delta H_{2,\phi}^{(\rm LO)} &= -\frac{g m_{V,\psi} }{2\pi R} \sum_{k = -\Lambda R}^{\Lambda R} \frac{k^2}{R^2} \frac{\Theta(2\omega_k-E_{\rm max})}{\omega_k^3}\int dx\, \phi\,.
\end{align}
These terms combine to give the full leading order  effective Hamiltonian with the fermion mass renormalization (dropping the identity contribution)
\begin{align}
H_{\rm eff,\,  m_\psi}^{(\rm LO)} = H_0 + V + \Delta V + \mathcal{C}^{\rm(LO)}_{\phi^2}  \int dx\, :\!\phi^2\!:+\ \mathcal{C}_\phi^{(\rm LO)} \int dx\, \phi
\end{align}
with $\mathcal{C}^{\rm(LO)}_{\phi^2}$ defined in \eq{eq:phi2lo} and 
\begin{align}
\mathcal{C}_\phi^{\rm (LO)} =  \frac{g m_{V,\psi} }{2\pi R} \sum_{k}\frac{k^2}{R^2} \frac{\Theta(E_{\rm max}-2\omega_k)}{\omega_k^3}
\end{align}
which is now finite for $\Lambda \rightarrow \infty$. Finally we can add in the NLO corrections to this effective Hamiltonian to improve convergence, resulting in the expression for the NLO effective Hamiltonian with fermion mass renormalization 
\begin{align}
\begin{split}
\label{eq:heff-final-renom}
H_{\rm eff,\,  m_\psi}^{(\rm NLO)} =&\ \left(1 + \mathcal{C}_{\mathbbm{1},1}^{\rm (NLO)}+ \mathcal{C}_{\mathbbm{1},2}^{\rm (NLO)}\right)H_0 + V + \Delta V + \mathcal{C}^{\rm(LO)}_{\phi^2}  \int dx\, :\!\phi^2\!:+\ \mathcal{C}_\phi^{(\rm LO)} \int dx\, \phi\\
&+ \mathcal{C}^{\rm (NLO)}_{\phi^2,\, 1} \left\{H_0, \int dx\, :\!\phi^2\!:\right\} +  \mathcal{C}^{\rm (NLO)}_{\phi^2,\, 2} \left[H_0, \int dx\, :\!\phi^2\!:\right]\\
&+ \mathcal{C}^{\rm (NLO)}_{\phi,\, 1} \left\{H_0, \int dx\, \phi\right\} +  \mathcal{C}^{\rm (NLO)}_{\phi,\, 2} \left[H_0, \int dx\, \phi\right]
\end{split}
\end{align}
with the additional coefficients
\begin{align}
\mathcal{C}_{\mathbbm{1},2}^{\rm (NLO)} &= - m_{V,\psi}^2 \frac{R}{E_{\rm max}}\label{eq:firstbla}\\
\mathcal{C}^{\rm (NLO)}_{\phi,\, 1} &= -\frac{g m_{V,\psi}}{2\pi E_{\rm max}}\\
\mathcal{C}^{\rm (NLO)}_{\phi,\, 2} &= -\frac{g m_{V,\psi}}{8\pi R}\sum_{k} \frac{k^2}{R^2} \frac{\Theta(2\omega_k - E_{\rm max})}{\omega^4_k}.\label{eq:cpihnlo}
\end{align}
Technically, $m_{V,\psi}$ introduces an additional scale in the system, which must remain compatible with the $1/E_{\max}$ low-energy expansion \eq{eq:LEexp}. From \eq{eq:firstbla}
this in turn requires $m_{V,\psi} \lesssim \sqrt{E_{\rm max}/R}$.

\section{Numerical results}
\label{sec:results}

We now present our numerical results for Yukawa theory, which we approach from the free Hamiltonian $H_0$ and its eigenstates in the Fock basis representation \eq{eq:focku}, truncated at energies below $E_{\rm max}$. We use  the effective Hamiltonians presented in Sec.~\ref{sec:Heff} and~\ref{sec:massRG} and  evaluate their eigenvalues and eigenvectors numerically  on this finite set of states. The result we obtain is nonperturbative in the Yukawa coupling $g$; its precision is limited instead by the finite reach in $E_{\rm max}$.

Fermion charge $Q$ and total momentum $P$ are conserved in this theory. We focus on the total momentum-zero sector, separated into the charge-0
and charge-1 sectors.\footnote{
Other sectors may also be of interest in future work, for example higher-charge sectors containing
fermion-fermion bound states, or boosted sectors, where finite-volume effects
can be reduced~\cite{Chen:2022zms}.} Charge conjugation acts on the creation operators as
\begin{align}
\mathcal C c_k^\dagger \mathcal C^{-1}=d_k^\dagger,\qquad
\mathcal C d_k^\dagger \mathcal C^{-1}=c_k^\dagger,\qquad
\mathcal C a_k^\dagger \mathcal C^{-1}=a_k^\dagger,
\end{align}
with $\mathcal C|0\rangle=|0\rangle$. The charge-0 states can then be further separated according to their charge-conjugation eigenvalue $C=\pm1$, which we denote by the
$C^+$ and $C^-$ sectors. The vacuum state is defined  as the lowest charge-0, total momentum zero state that is even under charge conjugation. We also employ a  definition of the physical fermion mass that holds nonperturbatively, as the difference between the lowest charge-1, total momentum zero state and the vacuum.  Unless otherwise stated, we work in units of the input fermion mass $m_\psi$. Our largest truncated Hilbert spaces contain $\approx 5\times 10^6$ states in a given symmetry sector.

\subsection{Convergence with $E_{\rm max}$}
\label{sec:convergence}

Hamiltonian truncation techniques rely on $E_{\rm max}$ convergence, which is improved by the effective approach outlined in the previous section. At finite volume, the discreteness of the free spectrum introduces additional structure in this convergence: varying $E_{\rm max}$ changes the truncated Hilbert space discontinuously as individual basis states cross the cutoff. The low-lying states we are interested in couple most directly to states near the cutoff with low particle number. A simple estimate of the characteristic energy spacing for these states comes from momentum-zero two-particle states with large equal and opposite momenta. At large momentum, increasing the magnitude of each momentum by one unit raises their total energy by approximately $2/R$. For our choice of volume $Lm_{\rm min}=2\pi Rm_{\rm min}=10$, with $m_{\rm min}=\min(m_\phi,m_\psi)$, this gives an energy spacing of $\Delta E/m_{\rm min}\simeq 2/(Rm_{\rm min})\simeq 1.26$. Since this structure is approximately periodic in $E_{\rm max}$, we refer to the position of $E_{\rm max}$ within one period as the \emph{cutoff phase}. We will use steps of $\Delta E_{\rm max}/m_{\rm min}=1.25$ to look at the $E_{\rm max}$ convergence, which approximately preserves the cutoff phase and tends to give smoother behavior. We verify and discuss this cutoff phase structure directly in Appendix~\ref{app:antrunc}.

\begin{figure}[h]
    \centering
\includegraphics[scale=0.75]{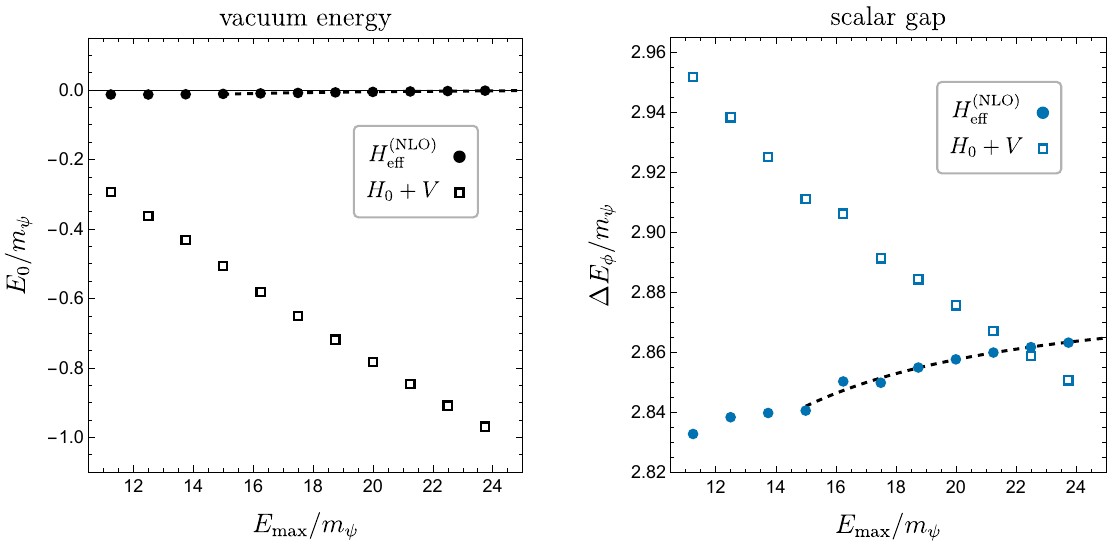}
    \caption{The $E_{\rm max}$ dependence of the vacuum energy and scalar gap for $m_\phi/m_\psi=3$ and $g/m_\phi=0.5$. Squares are the raw results  from $H_0+V$, while circles are the EFT-corrected results from \eq{eq:ident} and \eq{eq:HeffNLO}. Dashed lines denote $m_\psi^2/E_{\rm max}^2$ fits to the EFT-corrected data over the range~$E_{\rm max}/m_\psi\geq15$.}
\label{fig:convergence}
\end{figure}

A new feature of Yukawa theory, compared with previous implementations of HTET, is the presence of nontrivial UV divergences in both the vacuum energy and scalar self-energy, which make the corrections particularly important. In Fig.~\ref{fig:convergence} we compare predictions using the raw (uncorrected) Hamiltonian and the corrected effective Hamiltonian from \eq{eq:HeffNLO}, including the identity contribution from \eq{eq:ident} when studying the vacuum energy. The scalar gap is defined using the $C^+$ state in the charge-0 sector that connects continuously to the one-particle scalar state in the free theory, which we denote by $\Delta E_\phi$. For $m_\phi/m_\psi=3$ this is the second-lowest $C^+$ gap. The raw results do not converge with $E_{\rm max}$, consistent with the expected logarithmic divergence. Once the local counterterms and other EFT corrections are included, the results converge with the residual $1/E^2_{\rm max}$ scaling expected from power counting.

\begin{figure*}[h!]
    \centering
\includegraphics[scale=0.75]{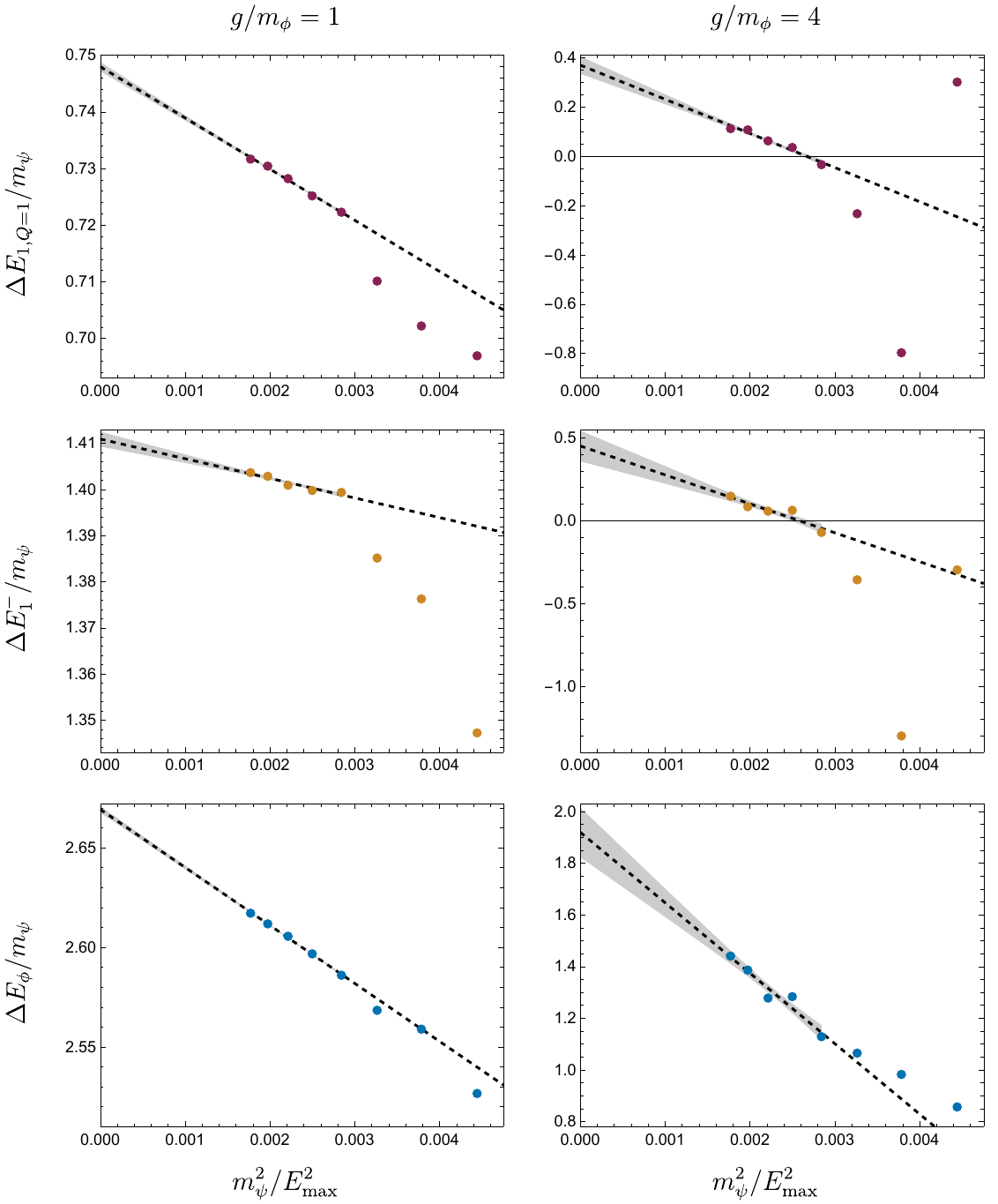}
    \caption{The remaining $E_{\rm max}$ dependence for three low-lying states with $m_\phi/m_\psi=3$, at $g/m_\phi=1$ (left) and $g/m_\phi=4$ (right). From top to bottom: the physical fermion mass ($Q = 1$ sector), the lowest $C^-$ gap, and the $C^+$ gap associated with the one-particle scalar state (both $Q = 0$ sector). Dashed lines are fits in $m_\psi^2/E_{\rm max}^2$ to the data with $E_{\rm max}/m_\psi\geq 18.75$, extrapolated to $E_{\rm max} \rightarrow \infty$, and the gray bands are the $1\sigma$ error bars, defined by $\sigma(x) = \sqrt{\sigma_a^2 + x^2\,\sigma_b^2 + 2x\,\sigma_{ab}\,}$, with $x=m_\psi^2/E_{\rm max}^2$ and $\sigma_{a,b,ab}$ the entries of the covariance matrix in the linear $a+bx$ fit.}
    \label{fig:scaling}
\end{figure*}

In Fig.~\ref{fig:scaling} we show the remaining $E_{\rm max}$ dependence for the full effective Hamiltonian~$H_{\rm eff}^{\rm (NLO)}$, for three representative low-lying states:  the physical fermion mass, the lowest $C^-$ gap in the $Q=0$ sector, and the $C^+$ gap associated with the one-particle scalar state. We show results for two representative couplings $g/m_\phi=1,4$. Our effective field theory implementation requires the IR scales of the theory to be  much smaller  than $E_{\rm max}$ and (from the power counting of Sec.~\ref{sec:EFT})  the leading remaining truncation errors scale as $1/E_{\rm max}^2$. We highlight this by showing the energies as functions of $m_\psi^2/E_{\rm max}^2$, so that we can linearly extrapolate the high-$E_{\rm max}$ points to $E_{\rm max}\rightarrow\infty$. In the figure the extrapolation is associated with the dashed lines, and the gray bands show the error.

At $g/m_\phi=1$ all three states show a clear $1/E_{\rm max}^2$ scaling. At $g/m_\phi=4$, where the theory is already very strongly coupled, deviations from the asymptotic scaling are larger at lower $E_{\rm max}$, as expected from the increased cutoff dependence at strong coupling; yet the extrapolation errors remain relatively low, roughly 5-10\%. The lowest $C^-$ state crosses below the $C^+$ vacuum for some values of $E_{\rm max}$, but the crossing disappears as the cutoff is increased. A similar vacuum crossing was observed in Ref.~\cite{Brooks:1983sb}, where it moved to larger coupling as the volume was increased, and was later found to disappear entirely at sufficiently large cutoff in Ref.~\cite{Pauli:1985ps}, as we find here.

\subsection{Spectrum and bound states}
\label{sec:spectrum}

Now that we have established convergence with $E_{\rm max}$, we look at the spectrum of the theory. At small coupling or in the large-$m_\phi$ regime we can compare with the results of Sec.~\ref{sec:an}; in the remaining parameter space, our results are genuinely new. The binding energy, which we study in Sec.~\ref{sec:BEs}, is defined relative to the physical two-particle threshold $2m_{\psi,\rm phys}$. To compare with these analytic results, we keep $m_{\psi,\rm phys}$ fixed, by tuning $m_{V,\psi}$ using the procedure of Sec.~\ref{sec:massRG}, as the coupling is varied. The tuned mass must remain within the regime of validity of the $1/E_{\rm max}$ expansion discussed below~\eq{eq:cpihnlo}. We consider separately the regimes $m_\phi>m_\psi$ and $m_\phi<m_\psi$.

\subsubsection{$m_\phi > m_\psi$}

\begin{figure*}[h!]
    \centering
\includegraphics[scale=0.74]{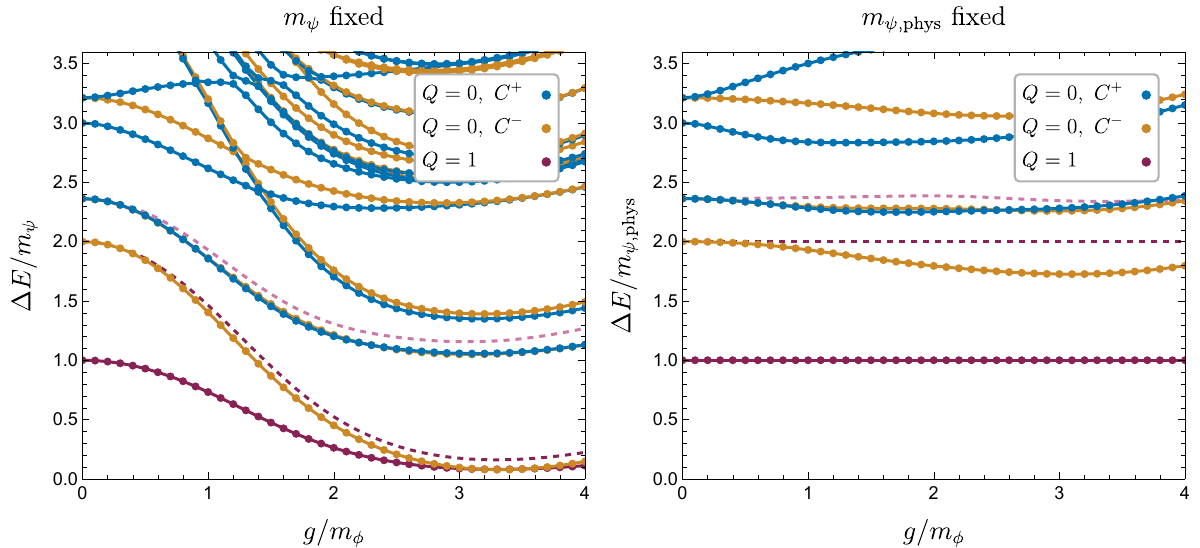}
    \caption{Spectrum of the lowest $P = 0$ states in the $Q = 0$ and $Q = 1$ sectors for $E_{\rm max}/m_\psi = 23.75$.
     In the left plot the input fermion mass $m_\psi$ is kept fixed, while on the right $m_{V,\psi}$ is numerically tuned so that $m_{\psi, \rm phys}$ is fixed. The lowest $Q = 1$,  $P = 0$ gap, which we identify with the physical fermion mass, is shown in purple. The remaining states are in the $Q = 0$, $P = 0$ sector, separated by their charge conjugation eigenvalue, $C^+$ (blue) or $C^-$ (orange). The two-particle threshold $2 m_{\psi, \rm phys}$ is shown as a purple dashed line, while twice the lowest $Q=1$, $P=1$ gap is shown as a pink dashed line. }
    \label{fig:spectrum}
\end{figure*}

In Fig.~\ref{fig:spectrum}, we show the energy gaps as a function of the dimensionless coupling $g/m_\phi$ in the momentum-zero sector, for fixed input $m_\phi$. In the left panel we also keep the input value of $m_\psi$ fixed, with $m_\psi = m_\phi/3$, while in the right panel we tune $m_{V,\psi}$ so that the physical fermion mass $m_{\psi,\rm phys}$ remains fixed to $m_\phi/3$.  These are two different slices of the theory, and the second choice allows for a more direct comparison with the limiting cases of Sec.~\ref{sec:an}.

We identify the lowest charge-1 gap as the physical fermion mass, and we separate the charge-0 states into $C^+$ and $C^-$ sectors. Without fermion mass renormalization, we see a consistent shift downward in all of the energy levels, as well as a  pairing of $C^+$ and $C^-$ states as the coupling is increased. After tuning $m_{V,\psi}$, the physical fermion mass remains constant and the downward trend is reduced, while the generic pairing of $C^+$ and $C^-$ states persists. In this regime, the most interesting states are the lowest-lying ones: the lightest charge-1 state, corresponding to the physical single-fermion state (shown in purple),  the lowest $C^-$ state (shown in orange), and the next-lowest $C\pm$ states in blue/orange. The low-lying charge-0 states include candidate bound states, as we now discuss.

Identifying bound states at finite volume and strong coupling is subtle, as there is no two-particle continuum to compare with, and because the very definition of bound state can be ambiguous. We instead compare potential fermion-antifermion bound states with the two-particle threshold, defined as twice the physical fermion mass, $2m_{\psi,\mathrm{phys}}$. For instance, in the spectrum in Fig.~\ref{fig:spectrum}, a candidate bound state is the lowest $C^-$ level,\footnote{Note that at finite volume in the free theory, the lowest fermion-antifermion state $c_0^\dagger d_0^\dagger \ket{0}$ is necessarily
$C^-$, as also supported by the discussion  of Sec.~\ref{sec:an}. Only with relative momentum between the two particles are there both charge-conjugation even and odd combinations.\label{ftnt:C}} which connects in the free theory to a fermion-antifermion state with zero relative momentum. The binding energy of this state is
\begin{align} \label{eq:bE}
E_{\rm B}^1 = 2m_{\psi, \rm phys} - \Delta E^-_{1}.
\end{align}
Similarly,  a candidate second bound state would correspond
-- at finite volume and in the free theory -- 
to a state with nonzero relative momentum between the two particles, with either charge conjugation eigenvalue (see footnote~\ref{ftnt:C}).
Its binding energy is defined as $E_{\rm B}^{2,\pm} = 2m_{\psi, \rm phys} - \Delta E^\pm_{2}$. Higher bound states can be found in a similar way and emerge from states that in the free theory have non-vanishing relative momentum. Since momentum is discrete at finite volume, these states already lie above twice the fermion mass. An alternative criterion for identifying the existence of a bound state would be to compare the energy of a state with the energy of its constituents carrying the same momenta. However, at finite volume an attractive interaction can lower a  state below the energy of the corresponding noninteracting two-particle level, so this comparison alone does not distinguish a bound state from an attractive scattering state~\cite{Sasaki:2006jn}. As~$L\to\infty$, the finite-volume momenta become continuous and the two prescriptions coincide, but at finite volume we use $2m_{\psi,\rm phys}$ as the reference.

In the right panel of \figref{fig:spectrum}, we can compare the energies of the first three candidate levels (the low-lying orange and blue curves) with twice the physical fermion mass (dashed purple line), see \eq{eq:bE}. The lightest $C^-$ state is always below the two-fermion threshold and we identify it with a bound state. The second and third states, on the other hand, remain above the two-fermion threshold for the parameters we study, so we do not identify them with higher bound states. Had we instead compared these states with the finite-volume energy of two fermions each with one unit of momentum (dashed pink line), both states would lie below this energy at low coupling. This would identify both as additional bound states in this regime, even though the non-relativistic analysis predicts only one bound state at weak coupling, and in the infinite-volume $m_\phi\to\infty$ spectrum the second one appears only for $g^2/m_\phi^2>\pi$.

\subsubsection{$m_\phi < m_\psi$}\label{sec:heavyf}

In the region of parameter space where the fermion is heavier than the mediating scalar, a second bound state can appear at moderate coupling, as shown in~\eq{eq:secondBS}. Here we study the case $m_\psi/m_\phi=3$, for which ~\eq{eq:secondBS} predicts the onset of a second bound state at $g/m_\phi\gtrsim1$. This is already at the edge of the regime where the perturbative analysis is controlled, so we use Hamiltonian truncation to study what happens at larger coupling.

\begin{figure*}[h!]
    \centering
\includegraphics[scale=0.65]{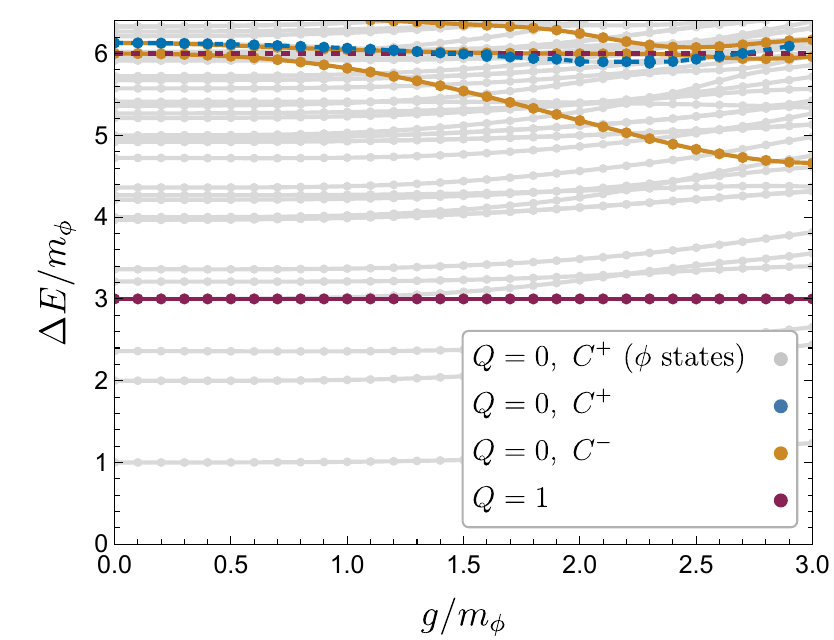}
    \caption{Spectrum of the lowest $P=0$, $Q=0$ gaps, together with the lowest $P=0$, $Q=1$ gap, for $m_\psi/m_\phi=3$, $E_{\rm max}/m_\phi=24$, and $Lm_\phi=10$, with $m_{\psi,\rm phys}$ held fixed. The $C^-$ states are shown in orange. Fermionic $C^+$ states are identified using significant ($\geq40\%$) overlap with the free-theory fermionic subspace together with continuity in energy as the coupling is varied, and are shown in blue, while the remaining $C^+$ states are shown in light gray.}
    \label{spectrumRG-heavyfer}
\end{figure*}
In Fig.~\ref{spectrumRG-heavyfer}, we show the low-lying states of the theory in the momentum-zero, charge-0 sector, as well as the fermion gap (the lowest momentum-zero, charge-1 state). We use the EFT-corrected Hamiltonian of~\eq{eq:heff-final-renom} and plot its eigenvalues as a function of increasing coupling, with cutoff $E_{\rm max}/m_\phi=24$ and volume $Lm_\phi=10$ -- notice that these are expressed in units of the scalar rather than the fermion mass. The physical fermion mass is held fixed along this scan. For this choice of masses, the lowest fermionic states lie among many multi-scalar states.\footnote{At finite volume, distinguishing a bound state from an attractive scattering level directly from the Hamiltonian spectrum can be ambiguous. Their different $L$-dependence can be used to distinguish the two~\cite{Luscher:1985dn,Luscher:1986pf}; we leave a systematic study of this volume dependence for future work.}
The $C^-$ states (shown in orange) can be identified using their charge-conjugation eigenvalue. The fermionic $C^+$ states (shown in blue) can mix with the nearby scalar states (shown in light gray), so instead we identify them using both non-negligible overlap with the free-theory fermionic subspace and continuity in energy as the coupling is varied.

\begin{figure*}[h!]
    \centering
\includegraphics[scale=0.65]{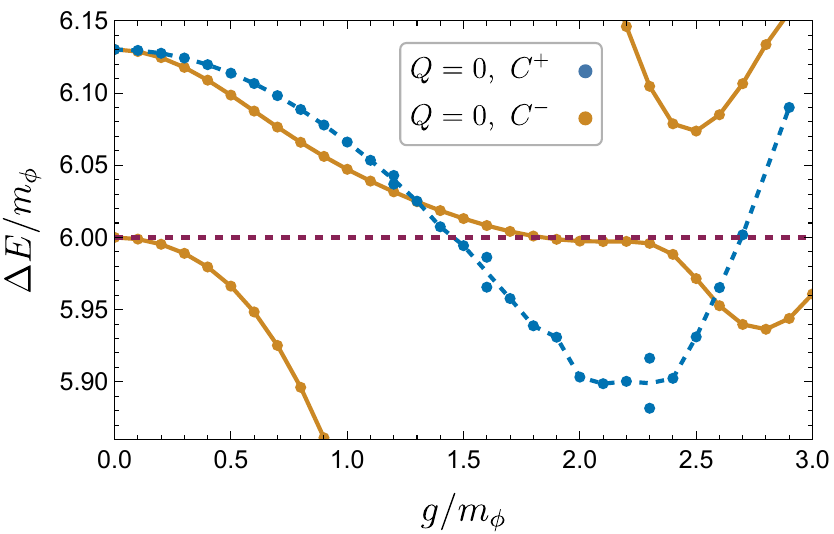}
    \caption{Close-up of the low-lying $P=0$, $Q=0$ states for the same parameters as Fig.~\ref{spectrumRG-heavyfer}. The $C^-$ states shown in orange are clearly distinguishable, while the lowest $C^+$ state shown in blue mixes with the nearby scalar states. We identify it using the fermionic overlap and energy-continuity criteria described in the text. Where its fermionic weight is distributed among nearby eigenstates, the individual states are shown and the blue dashed line connects their average energy as a guide to the eye. The dashed purple line shows the two-fermion threshold $2m_{\psi,\rm phys}$.}
\label{fig:twoboundstates}
\end{figure*}

A  close-up of the low-lying fermionic states is shown in Fig.~\ref{fig:twoboundstates}, where we compare their energies with twice the physical fermion mass. The lowest $C^-$ state remains below the two-fermion threshold throughout the range shown, consistent with the perturbative analysis of Sec.~\ref{sec:NRPTs}. As the coupling is increased, an additional $C^+$ fermionic state moves below the two-fermion threshold around $g/m_\phi\sim1.5$, in qualitative agreement with~\eq{eq:secondBS}, and remains clearly distinguishable over an intermediate range of coupling. At larger coupling, it becomes difficult to unambiguously follow a single $C^+$ state, since its fermionic weight can be distributed over many nearby states in the same sector. However, in this regime the value of $m_{V,\psi}$ required to keep the physical fermion mass fixed approaches the edge of the EFT validity discussed below~\eq{eq:cpihnlo}, so we do not draw conclusions about the fate of this state beyond the range shown.

\subsection{Binding energy}\label{sec:BEs}

The results of the previous section are in good qualitative agreement with the analytic expectations of Sec.~\ref{sec:an}. We now turn to a more quantitative study of the binding energy, comparing with finite-volume perturbation theory and the $m_\phi\to\infty$ Thirring/Sine-Gordon limit. For individual low-lying eigenvalues, the power counting of Sec.~\ref{sec:EFT} predicts a relative truncation error of
\begin{equation}
O\left(\max(g^2,m_\psi^2,m_\phi^2)/E_{\rm max}^2\right)\sim 10^{-2}   \,.\label{eq:errores}
\end{equation}
This residual cutoff dependence is small enough that the energy gaps show the clean scaling seen in Sec.~\ref{sec:convergence}. The binding energy, however, is the difference between the two-particle threshold and a nearby gap, Eq.~\eqref{eq:bE}. For some values of $g$ these two energies are very close, cutoff phase effects that are small on the scale of either energy  become relatively important in their difference. This is particularly visible at weak coupling, where the leading finite-volume result is $E_B\sim g^2/(m_\phi^2L)$.

The cutoff phase dependence appears as oscillations in the binding energy as $E_{\rm max}$ is varied. As described in Sec.~\ref{sec:convergence}, we chose our spacing $\Delta E_{\rm max}/m_{\rm min}=1.25$ so that successive points sample approximately the same cutoff phase, giving a smooth sequence in $E_{\rm max}$. In Appendix~\ref{app:antrunc}, we study this effect directly comparing with finite-volume perturbation theory with the same energy truncation as in the numerical calculation. There we find that shifted sets of cutoffs at approximately fixed cutoff phase are individually smooth, while the full finite-$E_{\rm max}$ result oscillates strongly as states enter the truncated basis. Averaging over the cutoff phase substantially reduces these fluctuations and approaches the infinite-$E_{\rm max}$ result much more smoothly. Similar cutoff effects, and the use of averaging/smoothing to reduce them, have been discussed also in~\cite{Rychkov:2014eea,Lencses:2015bpa,Elias-Miro:2017tup,Rutter:2018aog}.

\begin{figure*}[h!]
    \centering
\includegraphics[scale=0.85]{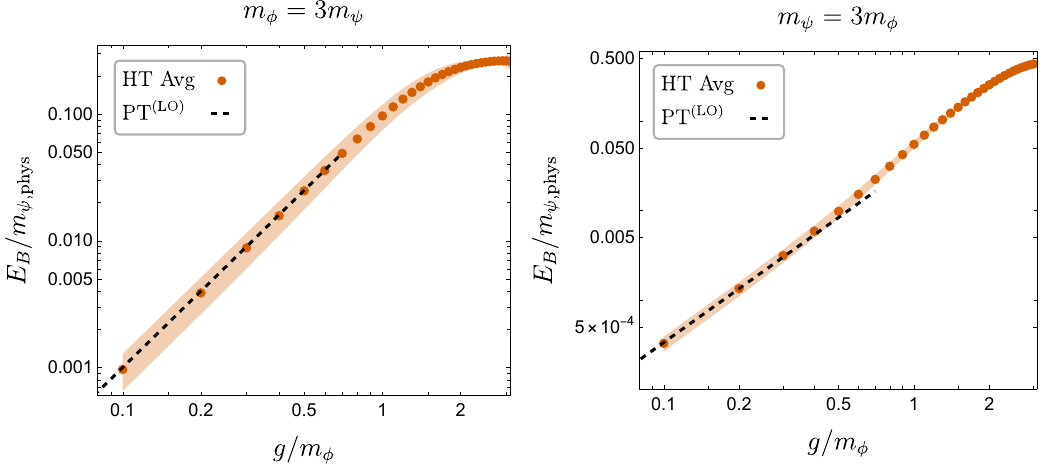}
   \caption{Binding energy of the lowest $C^-$ state as a function of $g/m_\phi$ in the two mass regimes of Sec.~\ref{sec:spectrum}. \textbf{Left:} $m_\phi/m_\psi=3$ and $Lm_\psi=10$. \textbf{Right:} $m_\psi/m_\phi=3$ and $Lm_\phi=10$. In both panels, $m_{\psi,\rm phys}$ is held fixed. The orange points show the Hamiltonian truncation result, averaged over five equally spaced cutoff phases spanning one period of the cutoff oscillations. For the left panel, these are $E_{\rm max}/m_{\psi,\rm phys}=22.75,\,23,\,23.25,\,23.5,\,23.75$, and for the right they are $E_{\rm max}/m_{\phi}=23,\,23.25,\,23.5,\,23.75,\, 24$. The shaded bands show the root-mean-square spread of the five cutoff-phase values about their mean. The black dashed lines show the leading finite-volume perturbative prediction, $E_B/m_{\psi,\rm phys}=(g/m_\phi)^2/(Lm_{\psi,\rm phys})$, evaluated for the parameters of each panel.}
    \label{fig:BEAvg}
\end{figure*}
For our calculations of the binding energy of the lowest $C^-$ state in both sectors, shown in Fig.~\ref{fig:BEAvg}, we therefore average over cutoff phase shifts. At each value of the coupling, we average the binding energy over five values of $E_{\rm max}$ separated by $\delta E_{\rm max}/m_{\rm min}=0.25$, which together span approximately one full period of the cutoff oscillations $\Delta E_{\rm max}/m_{\rm min} = 1.25$. 
The shaded bands in Fig.~\ref{fig:BEAvg} show the root-mean-square spread of these five values about their mean, which we use as a measure of the residual cutoff-phase dependence. As shown in Appendix~\ref{app:antrunc}, this averaging substantially reduces the cutoff-phase dependence and gives a much smoother approach to the infinite-$E_{\rm max}$ result. In Fig.~\ref{fig:BEAvg}, we show the resulting phase-averaged binding energy for $m_\phi/m_\psi = 3$ and $m_\psi/m_\phi = 3$. In both mass regimes, the binding energy agrees with the leading finite-volume perturbative prediction at weak coupling. In the $m_\phi<m_\psi$ regime, the departure from the leading-order result at somewhat larger coupling is consistent with higher-order perturbative corrections.


In the limit $m_\phi \rightarrow \infty$ with $g/m_\phi$ fixed, Yukawa theory approaches the massive Thirring model and its dual Sine-Gordon. In this limit, the lowest $C^-$ state becomes the first Sine-Gordon breather \eq{eq:breathers}, with the soliton mass identified with the physical fermion mass~$M_s=m_{\psi,\rm phys}$. Here we make a first attempt to approach this limit directly from Yukawa theory by considering $m_\phi/m_\psi=2$, $3$, and $4$.

\begin{figure*}[h!]
    \centering
\includegraphics[scale=0.7]{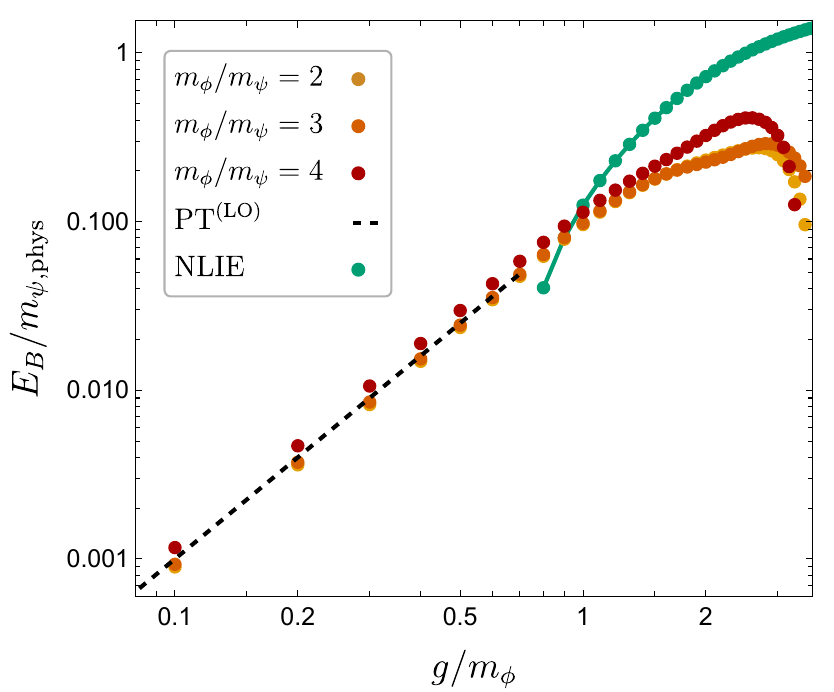}
    \caption{Binding energy as a function of $g/m_\phi$ for $m_\phi/m_\psi=2,\, 3,\, 4$ at $Lm_{\psi}=10$, with $m_{\psi,\rm phys}$ held fixed. Each  point is obtained by extrapolating five fixed-cutoff-phase sequences to $E_{\rm max}\to\infty$ using fits in $1/E_{\rm max}^2$ and averaging the extrapolated values. The largest cutoffs used are $E_{\rm max}/m_{\psi, \rm phys}=21.25$ for $m_\phi/m_{\psi}=2$, $E_{\rm max}/m_{\psi, \rm phys}=23.75$ for~$m_\phi/m_{\psi}=3$, and $E_{\rm max}/m_{\psi, \rm phys}=25$ for $m_\phi/m_{\psi}=4$. The black dashed line shows the leading finite-volume perturbative prediction, while the green points show the finite-volume Sine-Gordon prediction obtained from the NLIE described in 
Appendix~\ref{app:NLIE}.}
    \label{fig:SG}
\end{figure*}
As $m_\phi$ is increased, maintaining the hierarchy between the physical scales and the cutoff requires correspondingly larger values of $E_{\rm max}$, while the rapid growth of the truncated basis limits how far the cutoff can be pushed. The EFT corrections calculated in Sec.~\ref{sec:match} allow us to extrapolate $E_{\rm max}\rightarrow \infty$ using the predicted scaling, making it possible to control the cutoff dependence at the values of $E_{\rm max}$ that are computationally accessible. Obtaining a reliable prediction for the binding energy also requires averaging over the residual cutoff-phase dependence. For each of five fixed-cutoff-phase sequences, we compute the binding energy at five values of $E_{\rm max}$, tuning $m_{V,\psi}$ at each point to keep $m_{\psi,\rm phys}$ fixed. We then extrapolate each sequence to $E_{\rm max}\rightarrow\infty$ using fits in $1/E_{\rm max}^2$ and average the five extrapolated values. Averaging over the five cutoff phases before extrapolating gives very similar results.

In Fig.~\ref{fig:SG}, we compare the binding energy at these three scalar masses with the finite-volume perturbative result at weak coupling and the finite-volume Sine-Gordon prediction obtained from the NLIE of Appendix~\ref{app:NLIE}. At weak coupling, all three Yukawa results agree well with perturbation theory. At larger coupling, increasing $m_\phi$ moves the binding energy toward the NLIE prediction, with a clearer change between $m_\phi/m_\psi=3$ and $4$, although the $m_\phi/m_\psi=4$ result still does not reproduce the Sine-Gordon curve. A quantitative comparison with the Sine-Gordon limit is complicated by several different effects. At finite $m_\phi$ there are corrections to the leading large-$m_\phi$ matching onto the Thirring model, while increasing $m_\phi$ also reduces the accessible hierarchy $E_{\rm max}/m_\phi$, making the extrapolation more sensitive to higher-order cutoff corrections. The latter can be systematically improved by extending the effective Hamiltonian to higher order in $1/E_{\rm max}$. Finally, at fixed volume the single-source NLIE used here breaks down at weak coupling; extending the NLIE or going to larger volume would allow the comparison to be pushed further into this regime. We leave a systematic study of this~$m_\phi\to\infty$ limit to future work.

\section{Conclusions}
\label{sec:conclusions}

In this article we have developed a Hamiltonian truncation technique to study the spectrum of two-dimensional Yukawa theory across several regions of parameter space. Particular emphasis was put on the analysis of bound states, for which analytic predictions exist in the perturbative or $m_\phi\to\infty$ regimes. At weak coupling we identify a bound state in both mass regimes we study. This state connects to the fermion-antifermion pair at rest in the free theory, and its binding energy agrees well with finite-volume perturbation theory. Away from weak coupling, we follow the spectrum nonperturbatively into regions where no general analytic solution is available. We track how the first bound-state energy evolves across couplings, for different values of the scalar and fermion masses. In the regime of large fermion mass we also see the emergence of a second bound state at nonperturbative values of the coupling. This is qualitatively consistent with the non-relativistic analysis, which predicts the onset of a second bound state around $g/m_\phi\sim 1$ for $m_\psi/m_\phi=3$, at the edge of its regime of validity. We also study the large-scalar-mass regime as a first step toward the heavy-scalar limit, where Yukawa theory approaches the massive Thirring model, equivalent to Sine-Gordon. For these larger scalar masses, the binding energy agrees with finite-volume perturbation theory at weak coupling, and shows a promising qualitative trend toward the Sine-Gordon prediction at larger coupling.

These calculations required extending the HTET framework of Ref.~\cite{Cohen:2021erm} to a theory with both fermions and nontrivial UV divergences. After renormalizing the vacuum energy and scalar self-energy and including the effective Hamiltonian corrections derived using the diagrammatic rules of Appendix~\ref{app:Feyn}, the remaining dependence on $E_{\rm max}$ follows the $1/E_{\rm max}^2$ scaling expected from power counting. We also introduced a fermion mass term that can be tuned (numerically and nonperturbatively) as the coupling is varied, allowing us to keep the physical fermion mass fixed and measure binding energies relative to a fixed two-particle threshold. A quantitative comparison with finite-volume perturbation theory at weak coupling required a more precise determination of the binding energy. We find cutoff phase effects associated with the discrete finite-volume spectrum, which are small for the individual gaps but become significant for the binding energy. Averaging over the cutoff phase reduces this dependence, and we find good agreement with the finite-volume weak-coupling prediction. The perturbative calculation of Appendix~\ref{app:antrunc} independently reproduces the cutoff-phase structure. Similar behavior was already observed in Ref.~\cite{Demiray:2025zqh}, where these effects obscured the expected $1/E_{\rm max}^4$ scaling at smaller volume and were reduced at larger volume. A more detailed study of the cutoff phase effects, for instance in Fourier space, could help characterize their structure and develop more systematic ways of reducing them.

An important next step is to push the heavy-scalar calculation further toward the massive Thirring limit. The technical interest of this analysis lies in the fact that, while Yukawa theory has a relevant coupling, the Thirring model has a marginal four-fermion interaction. Convergence with the cutoff can be problematic for marginal interactions~\cite{Beria:2013hz}, and the Yukawa $\to$ Thirring approach could provide a route to treating marginal interactions, which is especially relevant in the context of attacking QCD~\cite{Fitzpatrick:2022dwq}. Such a development would require controlling the finite-$m_\phi$ corrections to the Yukawa/Thirring matching, as well as extending the effective Hamiltonian to higher order in the $1/E_{\rm max}$ expansion to reduce the truncation errors; this includes terms at $\mathcal{O}(g^3)$ and $\mathcal{O}(g^4)$ as predicted by the power counting in Sec.~\ref{sec:EFT}. As an additional check, we could simulate Sine-Gordon directly with Hamiltonian truncation at the same finite volume and compare the two calculations as the heavy-scalar limit is approached, including at weak coupling, where the NLIE as implemented in Appendix~\ref{app:NLIE} is not available. An intriguing feature of the Thirring/Sine-Gordon construction is the region $4\pi<\beta^2<8\pi$. In the Thirring model this corresponds to $-\pi<g^2/m_\phi^2<0$ and cannot be reached directly from Yukawa theory with real coupling. Yet, one can envisage starting from a non-Hermitian version of Yukawa theory with $g\in i\mathbb{R}$ and an antilinear PT symmetry that, when unbroken, would guarantee a real spectrum~\cite{Bender:2007nj}. Non-Hermitian theories have been studied before by Hamiltonian truncation in~\cite{Yurov:1989yu,Delouche:2023wsl,Delouche:2024yuo,Lencses:2023evr}. We leave all this to future work.

The methods developed here also provide a starting point for studying Gross-Neveu-Yukawa theories in equal-time Hamiltonian truncation, where scalar self-interactions lead to a richer set of infrared fixed points. The two-dimensional supersymmetric Gross-Neveu-Yukawa model has already been studied using lightcone conformal truncation, with supersymmetry built directly into the construction~\cite{Fitzpatrick:2019cif}. Starting from the generic Gross-Neveu-Yukawa theory in equal time we could test the proposed emergence of supersymmetry at the tricritical Ising fixed point~\cite{Fei:2016sgs}. The same framework could also probe the non-supersymmetric fixed point, for which a non-unitary CFT has been proposed in two dimensions~\cite{Nakayama:2022svf}.

\subsection*{Acknowledgments}
We thank  M.~Walters for important discussions, D.~Kosmopoulos for discussions and collaboration on a related project, and especially D. Wenzel for collaboration on the initial stages of this project.
F.R. is supported by the SNSF under grant no. 200021205016, while  O.D. is supported  by LEAD project 200021E205315. F.R. and O.D. also  thank M.~Cirelli for discussions and 
 acknowledge financial support from the 4EU+ Alliance within the framework of the DaCoSMiG project.

\appendix

\section{Finite-volume Sine-Gordon spectrum from a NLIE}\label{app:NLIE}

This Appendix collects the equations behind the finite-volume Sine-Gordon curve of Figs.~\ref{fig:SGNR} and \ref{fig:SG}, the numerical procedure used to solve them, and the estimates that delimit its regime of validity in~\figref{fig:validity}. The $1+1$ dimensional kinematics are expressed in terms of the rapidity $\theta$, itself 
related to energy and momentum by $ E = M\cosh\theta$, $p = M\sinh\theta$ for a particle of mass $M$. Bound states appear at imaginary rapidity: for instance
the infinite-volume first breather  $M_1 = 2m_\psi\sin(\pi\xi/2)$ can be thought of as a
 soliton-antisoliton bound state with imaginary relative rapidities $\pm\theta_\infty=\pm i \pi(1-\xi)/2$ and $\theta_\infty\to \theta_\infty +i\pi$ in the crossed channel. Throughout this Appendix, $m_\psi$ denotes the soliton mass,
$m_\psi=m_{\psi,\rm phys}=M_s$.

In finite volume, the Sine-Gordon states considered here can be encoded in a single \emph{counting function} $Z(\theta)$, by construction a monotonic function of rapidity.
A  generalization of the quantization of momenta of free particles in a box leads to a  nonlinear integral equation -- NLIE -- for $Z$~\cite{Destri:1992qk,Destri:1997yz,Feverati:1998dt},
\begin{equation}\label{eq:NLIE}
\begin{split}
Z(\theta)=\ell\sinh\theta+g_{\rm src}(\theta)
&+\int_{-\infty}^{\infty}\frac{dx}{i}\;G(\theta-x-i\eta)\,\ln\!\left[1+(-1)^\delta e^{iZ(x+i\eta)}\right]\\
&-\int_{-\infty}^{\infty}\frac{dx}{i}\;G(\theta-x+i\eta)\,\ln\!\left[1+(-1)^\delta e^{-iZ(x-i\eta)}\right],
\end{split}
\end{equation}
 with dimensionless volume $\ell\equiv m_\psi L$ and kernel
\begin{equation}\label{eq:NLIEkernel}
G(\theta)=\int_{-\infty}^{\infty}\frac{dk}{2\pi}\,e^{ik\theta}\,
\frac{\sinh\frac{\pi(\xi-1)k}{2}}{2\sinh\frac{\pi\xi k}{2}\,\cosh\frac{\pi k}{2}}\,,
\end{equation}
fixed by the phase of the exact soliton--soliton S-matrix~\cite{Zamolodchikov:1978xm}.
Here the first term on the right-hand side of \eq{eq:NLIE} is  the free-theory phase, the last two terms capture the phase picked up by scattering off real or virtual states, the sum has been converted into an integral and the integration contour deformed into the complex plane: $\eta\gtrsim 0$ shifts the integration contours off the real axis (results are $\eta$-independent), while $\delta\in\{0,1\}$ selects integer/half-integer quantization conditions (the vacuum/first breather have $\delta=0/1$ respectively). The source $g_{\rm src}(\theta)$ carries the information identifying the state, as an analytic continuation of the scattering phase from the associated infinite-volume rapidity (e.g. $\theta_\infty$ for the first breather). The solution for $Z$ determines the allowed rapidities. Summing the corresponding
single-particle energies $m_\psi \cosh\theta$ and converting the sum into a contour integral
gives

\begin{equation}\label{eq:NLIEenergy}
E(L)=E_{\rm src}-\frac{m_\psi}{2\pi}\int_{-\infty}^{\infty}dx\,\sinh x\;\mathcal{Q}(x)\,,\qquad
\mathcal{Q}(x)\equiv\frac{1}{i}\ln\frac{1+(-1)^\delta e^{iZ(x+i\eta)}}{1+(-1)^\delta e^{-iZ(x-i\eta)}}\,,
\end{equation}
where $E_{\rm src}$ is the explicit contribution of the sources. In principle, the branch of the logarithm in $\mathcal{Q}$ must be followed continuously along the contour, but when $Z$ is monotonic on the real axis it can be replaced by its principal value.

For the binding energy, we need the \emph{vacuum}, which has no sources ($g_{\rm src}=0$, $\delta=0$, $E_{\rm src}=0$, giving $E_0(L)$) and the \emph{first breather at rest}, with infinite-volume bound-state rapidity  $\theta_\infty\to i\pi(1+\xi)/2$, $\delta=1$, and associated source,
\begin{equation}\label{eq:B1source}
g_{\rm src,1}(\theta)=-i\ln\frac{\cos\frac{\pi\xi}{2}-i\sinh\theta}{\cos\frac{\pi\xi}{2}+i\sinh\theta}\,,\qquad
E_{\rm src}=2m_\psi\sin\frac{\pi\xi}{2}=M_1(\infty)\,,
\end{equation}
giving the first-breather state energy $E_1(L)$. The finite-volume breather mass and binding energy are then
\begin{equation}\label{eq:NLIEgap}
M_1(L)=E_{1}(L)-E_0(L)\,,\qquad E_B(L)=2m_\psi-M_1(L)\,.
\end{equation}
As $L\to\infty$ the integrals in \eq{eq:NLIEenergy} vanish exponentially and $M_1(L)\to M_1(\infty)$, as it should.

\paragraph{Breakdown at weak coupling.} While \eq{eq:NLIE} is in principle exact (there is no expansion in the volume or in the coupling), the single-source description \eq{eq:B1source} is not valid at arbitrarily weak coupling for fixed volume. Physically, this is related to the fact that, as $\xi\to1$, the binding energy vanishes and the size $L_B$ of the bound state grows; when $L_B\sim L$ the level is better described as a fermion--antifermion pair delocalized over the volume (the starting point of NRPT). In the NLIE, this appears through the ``counting'' function  $Z$ losing its built-in monotonicity and developing non-analytic structures in the complex $\theta$ plane~\cite{Destri:1997yz,Feverati:1998dt,Feverati:2000xa}.
The breakdown can be estimated by the slope of the counting function at the origin, neglecting the convolution integrals in \eq{eq:NLIE}, $Z'(0)\simeq\ell-2/\cos(\pi\xi/2)$ changes sign at
\begin{equation}\label{eq:gcmono}
\xi_c(\ell)=\frac{2}{\pi}\arccos\frac{2}{\ell}\,.
\end{equation}

Translated to the coupling through $g/m_\phi=\sqrt{\pi(1-\xi)/\xi}$, this gives $g_c/m_\phi\simeq0.68(1.65)$  at $\ell=10(3)$, the red dashed line in \figref{fig:SGNR}, below which we do not display NLIE results.
In principle, a modified source ansatz can account for the loss of monotonicity -- we do not attempt it here.

\paragraph{Numerical solution.}
In the regime where we trust the NLIE, we solve it by discretizing the rapidity on a uniform grid of $N$ points on $[-\Lambda,\Lambda]$ and represent
$Z$ on the two shifted contours $x\pm i\eta$,\footnote{The kernel has singularities at $\theta=\pm i\pi\xi$, forcing $|2\eta|<\pi\xi$, which becomes more stringent at strong coupling, where $\xi$ is
small.} on which the nonlinear terms of \eq{eq:NLIE} are
defined; since $Z$ is real on the real axis, $Z(x-i\eta)=\overline{Z(x+i\eta)}$, we can focus on the upper contour. Both integrals in \eq{eq:NLIE} are convolutions and we evaluate them as  products in Fourier space.  We use $\Lambda=14$, $N=2^{12}=4096$ and $\eta=0.10$ -- and  test stability under changes of these values. At every step in the iteration we define
\begin{equation}\label{eq:damped}
Z^{(n+1)}=Z^{(n)}+\alpha\left(\mathcal{F}[Z^{(n)}]-Z^{(n)}\right)\,,
\end{equation}
with $\alpha=0.4$ providing a mixture of the $n$-th iteration and the new one $\mathcal{F}[Z^{(n)}]$ in the right-hand side of \eq{eq:NLIE} (the ``undamped'' version with $\alpha=1$ does not converge). We stop when
~$\max_x|Z^{(n+1)}-Z^{(n)}|<10^{-9}$.

Deep in the region of validity  (for $g\gg g_c$ of the previous
subsection), the iteration started
from $Z^0=\ell\sinh\theta$ converges in a few tens of steps and the solution is unambiguous. 
Near the boundary of validity, instead, the iteration  develops competing fixed points. 

We select a solution  only when the following two criteria are satisfied: at each coupling, solutions from scratch with $\eta=0.10$ and $0.16$ agree to $10^{-6}$, \emph{and} they smoothly agree with a continuation from the previous
accepted point in $g$, in steps $\Delta(g/m_\phi)=0.02$.
We also verify a posteriori that 
 the analytic structure assumed in deriving
\eq{eq:NLIE} is the one actually realized: 
$Z$ remains monotonic
on the real axis and the singularities 
stay far from the integration contours, with 
the only one in the strip $0<\mathrm{Im}\,\theta<\pi$ being that associated with the source
\eq{eq:B1source}.

The implementation was validated  against the finite-volume NLIE
spectra of Ref.~\cite{Feverati:2000xa} for the vacuum, one-soliton and first-breather energies
tabulated there at $\xi=2/7$ and $2/9$, which we reproduced to all quoted digits for different volumes, down to $\ell=3$.

\section{Yukawa diagrammatic rules}\label{app:Feyn}

Here we present a set of diagrammatic rules used to compute the counterterms and effective
Hamiltonian corrections in Yukawa theory. These are old-fashioned (Hamiltonian) perturbation theory rules, where different time orderings correspond to different diagrams. Diagrams do not need to be individually covariant, although their sum is covariant in the untruncated theory. We work in the Fock basis \eq{eq:focku}. For a calculation at $n$th order in the interaction, the diagrammatic rules are:

\begin{itemize}
\item Include all possible diagrams with $n$ vertices, including disconnected ones. Different vertex orderings are treated as separate diagrams, with time flowing from right to left. Normal ordering forbids lines from beginning and ending at the same vertex.
\item Assign a spatial momentum label to all internal and external lines and sum over the allowed discrete internal momenta. Every internal line is on shell with positive energy, \begin{equation} k^\mu = \left(\omega_k,\frac{k}{R}\right) \end{equation}
with $\omega_k \rightarrow \omega_{\phi,k}$ for scalars. The arrows on fermion lines indicate fermion-number flow, which for an antifermion runs opposite to the direction of propagation. The label on a line is always the physical spatial momentum of the particle propagating along it, so an antifermion line labeled $k$ has numerator $\slashed{k}-m_\psi$ with
$k^\mu=(\omega_k,k/R)$. Momentum is conserved at each vertex.
\item For each external particle, include the following factors:
\begin{align}
\begin{split}
\textnormal{final state antifermion:}\quad{}
&\bm{v}_k\\
\textnormal{initial state antifermion:}\quad{}
&\bar{\bm{v}}_{\ell}\\
\textnormal{final state fermion:}\quad{}
&\bar{\bm{u}}_k\\
\textnormal{initial state fermion:}\quad{}
&\bm{u}_\ell \\
\textnormal{final state scalar:}\quad{}
&1\\
\textnormal{initial state scalar:}\quad{}
&1.
\end{split}
\end{align}
\item For each Yukawa vertex, include:
\begin{equation}
\includegraphics[scale=0.5,trim={0cm 0cm 0cm 0cm},clip,valign=c]{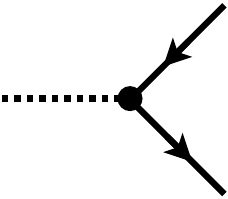} = \frac{g}{\sqrt{2\pi R}}.
\end{equation}
\item For each internal fermion, antifermion or scalar propagator, include:
\begin{align}
\includegraphics[scale=0.5,trim={0cm 0cm 0cm 0cm},clip,valign=c]{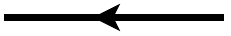} &=  \frac{\slashed{k}+m_\psi}{2\omega_k}\\
\includegraphics[scale=0.5,trim={0cm 0cm 0cm 0cm},clip,valign=c]{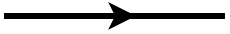} &=  \frac{\slashed{k}-m_\psi}{2\omega_k}\\
\includegraphics[scale=0.5,trim={0cm 0cm 0cm 0cm},clip,valign=c]{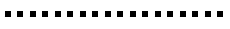} &= \frac{1}{2\omega_{\phi,k}}.
\end{align}
\item Multiply the spinor factors along each fermion line, following the fermion-number arrow backwards. If the line ends on an external fermion or antifermion, contract with the corresponding external spinor; for a closed fermion loop, take the trace. The signs associated with closed fermion loops are fixed by the ordering of the Fock basis, so no additional factors of $-1$ are needed.

\item For each intermediate state include an energy denominator:
\begin{align}
\frac{1}{E_f - E_\alpha + i \epsilon},
\end{align}
where $E_f$ is the final state energy and $E_\alpha$ is the energy of the intermediate state, including spectator particles. In the truncated theory, also include a factor:
\begin{align}
\Theta(E_{\rm max} - E_\alpha)
\end{align}
for each intermediate state, restricting the sum to states with energy less than or equal to $E_{\rm max}$.
\item For external states, include a factor of the form:
\begin{align}
\langle f |\phi^{(-)}_{p_1} \hat{\bar{\psi}}^{(-)}_{p_2}\hat{\psi}^{(-)}_{p_3}\cdots \phi^{(+)}_{q_1}\hat{\psi}^{(+)}_{q_2}\hat{\bar{\psi}}^{(+)}_{q_3}|i\rangle
\end{align}
corresponding to the correct initial and final state particles and their assigned momenta. The hatted components are defined as in \eq{eq:fermioncomps}, with the spinors stripped off.
\end{itemize}

With the inclusion of the fermion mass renormalization interaction in Sec.~\ref{sec:massRG}, we also have the additional vertex rule:
\begin{equation}
\includegraphics[scale=0.5,trim={0cm 0cm 0cm 0cm},clip,valign=c]{Figs/Yukawa-FermionVertex} = m_{V,\psi}.
\end{equation}

\section{Binding energy at finite truncation}\label{app:antrunc}

\begin{figure}[h]
    \centering
     \begin{subfigure}[t]{0.28\textwidth}
       \centering
    \raisebox{0.5cm}[0pt][0pt]{%
        \includegraphics[width=0.5\linewidth]{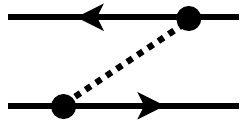}
    }
        \caption{Tree-level exchange.}
    \end{subfigure}
    \begin{subfigure}[t]{0.35\textwidth}
        \centering
        \raisebox{0.3cm}[0pt][0pt]{%
        \includegraphics[width=\linewidth]{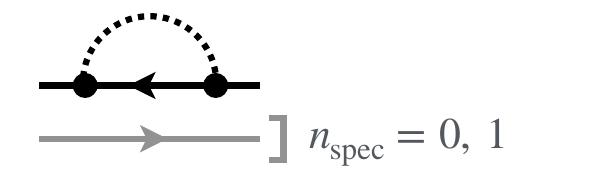}
    }
        \caption{One-loop $F$ term.}
        \label{fig:Fdiag}
    \end{subfigure}
    \begin{subfigure}[t]{0.35\textwidth}
        \centering
        \includegraphics[width=\linewidth]{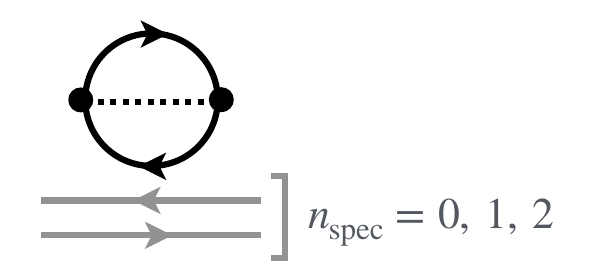}
        \caption{Two-loop $G$ term.}
    \end{subfigure}

 \caption{Second-order diagrams for the three external states. Gray lines denote spectator particles. For the $F$ term, $n_{\rm spec}=0,\,1$ for the one-fermion and fermion-antifermion states, respectively. For the $G$ term, $n_{\rm spec}=0,\,1,\,2$ for the vacuum, one-fermion, and fermion-antifermion states, respectively. Only one of the two vertex orderings is shown for the tree-level diagram, the second vertex ordering for the one-loop diagram is forbidden by Pauli exclusion.}
    \label{fig:BEdiagrams}
\end{figure}

In this appendix, we use old-fashioned perturbation theory to compute the leading contribution to the binding energy of the first bound state in Yukawa theory at finite volume. At weak coupling, the relevant states are the vacuum, the zero-momentum one-fermion state, and the fermion-antifermion state with both particles at zero momentum:
\begin{align}
\label{eq:BEstates}
\ket{0}, \quad{} c_0^\dagger\ket{0}, \quad{} c_0^\dagger d_0^\dagger \ket{0}.
\end{align}
We denote the free-theory energies of these states by $E_0$, $E_{1f}$ and $E_{2f}$, respectively. Since $V$ has no nonzero diagonal matrix elements, the leading correction to the energy of a state $\ket{i}$ is second order,
\begin{equation} \label{eq:2nd-corr}
E^{(2)}_i = \sum_{\ket{k}\neq \ket{i}}\frac{|\bra{k}V\ket{i}|^2}{E_i-E_k}\,,
\end{equation}

where $V$ is the Yukawa interaction and $E_i,\ E_k$ are the free-theory energies of $\ket{i}, \ket{k}$, respectively. The binding energy is obtained from twice the one-fermion gap minus the fermion-antifermion gap, so at this order
\begin{equation}
\label{eq:EB2def}
E_B^{(2)} = 2E_{1f}^{(2)} - E_{2f}^{(2)} - E_0^{(2)}.
\end{equation}
For the three external states above, the second-order corrections contain three diagrammatic structures, shown in Fig.~\ref{fig:BEdiagrams}: a tree-level scalar exchange, a one-loop fermion self-energy term (denoted by $F$) and a two-loop vacuum-energy term (denoted by $G$).

These contributions can be evaluated directly from the Yukawa interaction in Eq.~\eqref{eq:yuk}, or equivalently using the diagrammatic rules of Appendix~\ref{app:Feyn}. For the three states in Eq.~\eqref{eq:BEstates}, this gives

\begin{align}
    E_0^{(2)} =&\ \sum_{p,q} G(p,q)\,,\\
    E_{1f}^{(2)} =& \ \sum_{\substack{p,q\\ p\neq 0}} G(p,q)+ \sum_{\ell} F(\ell)\,,\label{eq:E1f}\\
    E_{2f}^{(2)} =& \ \sum_{\substack{p,q\\ p\neq 0,\ q\neq 0}} G(p,q) + 2\sum_\ell F(\ell)  - \frac{g^2}{2\pi R m_\phi^2},\label{eq:E2f}
\end{align}
where
\begin{align}
    F(\ell) \equiv&\ \frac{g^2}{8\pi R}\,\frac{m_\psi+\omega_\ell} {\omega_\ell\,\omega_{\phi,\ell} \left(m_\psi-\omega_\ell-\omega_{\phi,\ell}\right)}\,,\\
    G(p,q) \equiv&\ -\frac{g^2}{8\pi R}\,\frac{\omega_p\omega_q-m_\psi^2-pq/R^2}
{\omega_p\,\omega_q\,\omega_{\phi,p+q}
\left(\omega_p+\omega_q+\omega_{\phi,p+q}\right)}\, .
\end{align}
The function $G(p,q)$ is the summand of the vacuum-energy contribution in Eq.~\eqref{eq:vac}. The restrictions on the $G$ sums in Eq.~\eqref{eq:E1f} and Eq.~\eqref{eq:E2f} are from Pauli exclusion when there are zero-momentum fermions/antifermions already present in the external state.\footnote{Pauli exclusion need not be imposed on intermediate states, provided the same prescription is used consistently throughout the calculation, including the vacuum terms \cite{Feynman:1949hz}. For example, for the one-fermion state $c_0^\dagger\ket{0}$, if we did not impose Pauli exclusion, we would include the $p=0$ term in the $G$ sum in Eq.~\eqref{eq:E1f}. We would also need to include the other vertex ordering of the one-loop fermion self-energy diagram in Fig.~\ref{fig:Fdiag} (not shown), whose contribution \emph{exactly} cancels the $p=0$ part of $G$. Imposing Pauli exclusion removes both terms from the outset, giving the same $E_{1f}^{(2)}$. Note this equivalence holds only in the untruncated theory; at finite $E_{\max}$, we impose Pauli exclusion explicitly because our intermediate states are Fock states with Pauli exclusion built in.} For the fermion-antifermion state, either external particle can receive the self-energy correction in Fig.~\ref{fig:Fdiag}, giving the factor of two multiplying $F$ in Eq.~\eqref{eq:E2f}. The final term in $E_{2f}^{(2)}$ comes from the two vertex orderings of the tree-level scalar exchange.

This result is obtained without renormalization, so the $G$ sums associated with the vacuum-energy diagram are logarithmically divergent. Energy gaps are insensitive to this divergence, and so is the binding energy.\footnote{Note also that the vacuum, one-fermion, and fermion-antifermion states are insensitive to the scalar mass divergence at leading order.} We can see this explicitly by regulating the sums with the same UV cutoff $\Lambda$ introduced in Sec.~\ref{sec:CT}, so that $|p|,|q|\leq \Lambda R$. Substituting Eqs.~\eqref{eq:E1f} and \eqref{eq:E2f} into \eqref{eq:EB2def}, the $F$ terms cancel, and the $G$ terms give
\begin{align}
2\sum_{\substack{p,q\\p\neq0}}G(p,q)
-\sum_{\substack{p,q\\p\neq0,\ q\neq0}}G(p,q)
-\sum_{p,q}G(p,q)
=-G(0,0)=0,
\end{align}
where we used $G(p,q)=G(q,p)$. We are left with only the tree-level term
\begin{equation}
E_B^{(2)}
=\frac{g^2}{2\pi Rm_\phi^2}
=\frac{g^2}{m_\phi^2L},
\end{equation}
reproducing the leading small-volume term of Eq.~\eqref{eq:anapp}.

This calculation was performed in the untruncated theory. At finite $E_{\max}$, however, Hamiltonian truncation restricts the sum in Eq.~\eqref{eq:2nd-corr} to intermediate states $\ket{k}$ with $E_k\leq E_{\max}$. Accounting for the different intermediate states that contribute to each energy correction, the $\mathcal{O}(g^2)$ binding energy at finite $E_{\max}$ is
\begin{equation} \label{eq:g2coeff}
E_{B}^{(2)}(E_{\max}) 
= \sum_{p,q}\chi_G(p,q;E_{\max})G(p,q) +\sum_\ell \chi_F(\ell;E_{\max})F(\ell) +\frac{g^2}{2\pi Rm_\phi^2},
\end{equation}
where we define the cutoff weights
\begin{align}
\begin{split}
\chi_G(p,q;E_{\max})
={}&
2\,\Theta\!\left(E_{\max}
-\omega_p-\omega_q-\omega_{\phi,p+q}-m_\psi\right)
(1-\delta_{p,0})
\\
&-
\Theta\!\left(E_{\max}
-\omega_p-\omega_q-\omega_{\phi,p+q}-2m_\psi\right)
(1-\delta_{p,0})(1-\delta_{q,0})
\\
&-
\Theta\!\left(E_{\max}
-\omega_p-\omega_q-\omega_{\phi,p+q}\right),
\end{split}\\
\chi_F(\ell;E_{\max})
={}&
2\,\Theta\!\left(E_{\max}-\omega_\ell-\omega_{\phi,\ell}\right)
-
2\,\Theta\!\left(E_{\max}-\omega_\ell-\omega_{\phi,\ell}-m_\psi\right).
\end{align}
Each $\Theta$ function imposes the cutoff on the corresponding intermediate state in Eq.~\eqref{eq:2nd-corr}. The $m_\psi$ and $2m_\psi$ terms are the energies of the zero-momentum spectator fermions in the external states, while the Kronecker deltas enforce Pauli exclusion. At leading order in the EFT expansion of Sec.~\ref{sec:EFT}, the spectator energies are dropped from the cutoff conditions and the $F$ and $G$ terms cancel among themselves, leaving the untruncated result $E_B^{(2)}=g^2/(m_\phi^2L)$. The finite-$E_{\max}$ fluctuations arise from the state-dependent shifts of the truncation boundary.
\begin{figure*}[h!]
    \centering
\includegraphics[scale=0.85]{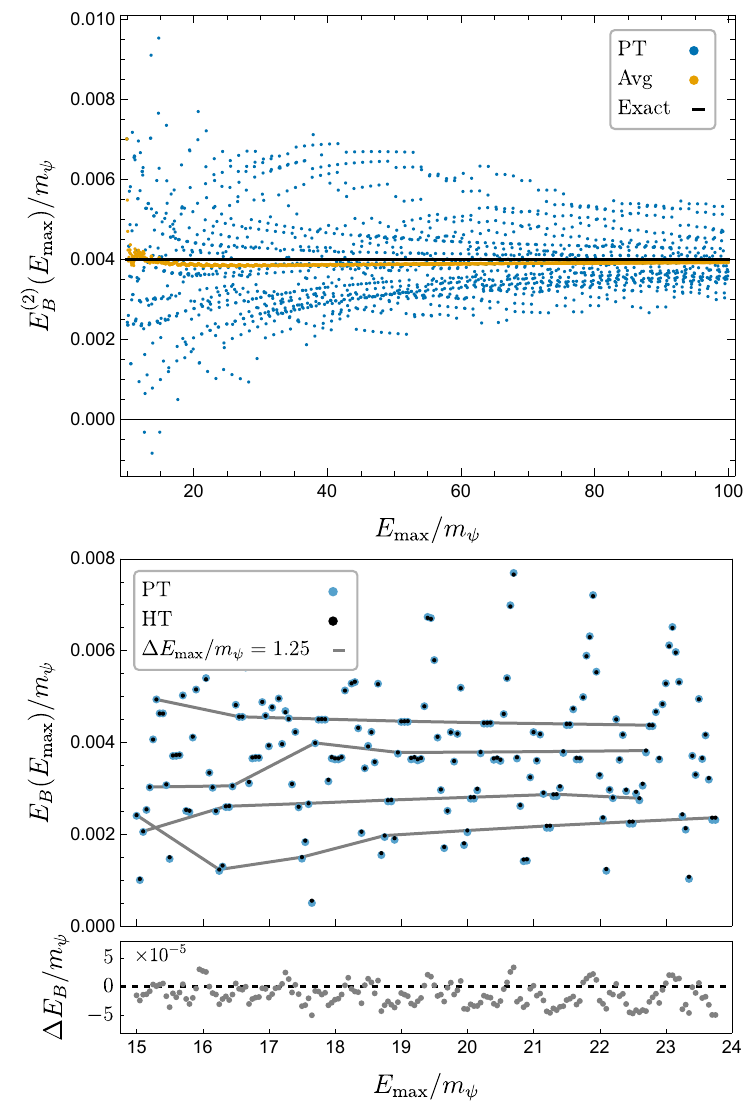}
 \caption{Binding energy at finite truncation, for $Lm_\psi=10$, $m_\phi/m_\psi=3$, and $g/m_\phi=0.2$. \textbf{Top:} The perturbative result over a wide range of $E_{\rm max}$, evaluated using a spacing $\Delta E_{\rm max}/m_\psi=0.05$. The yellow points show the cumulative average over the sampled cutoff values, while the black horizontal line shows the infinite-$E_{\rm max}$ result $E_B^{(2)}/m_\psi=0.004$. \textbf{Bottom:} Comparison of the perturbative result with Hamiltonian truncation using $H_{\rm eff}^{\rm NLO}$ over the accessible cutoff range. Gray lines connect shifted sets of cutoff values separated by $\Delta E_{\rm max}/m_\psi=1.25$, illustrating the smooth sequences discussed in Sec.~\ref{sec:convergence}. The lower panel shows the difference between the perturbative and Hamiltonian truncation results.}
    \label{fig:BEscatter}
\end{figure*}

We can now evaluate the finite-$E_{\max}$ expression \eqref{eq:g2coeff} numerically. The top panel of Fig.~\ref{fig:BEscatter} shows the perturbative result over a wide range of $E_{\max}$. Since the perturbative expression only requires finite sums, it can be evaluated at values of $E_{\max}$ far beyond those accessible to Hamiltonian truncation. The binding energy oscillates strongly as discrete intermediate states cross the cutoff, while the cumulative average approaches the infinite-$E_{\max}$ result much more smoothly. The bottom panel focuses on a range of $E_{\rm max}$ more accessible to truncation and compares the perturbative calculation at $g/m_\phi=0.2$ with Hamiltonian truncation using $H_{\rm eff}^{\rm NLO}$, with the difference between the two shown in the lower panel. The Hamiltonian truncation results closely follow the perturbative prediction point by point, including its cutoff-dependent fluctuations.

The detailed structure of these oscillations also provides a check of the cutoff spacing used in Sec.~\ref{sec:convergence}. There we estimated that at large momentum successive jumps should be separated by $\Delta E_{\rm max}/m_\psi\simeq2/(Rm_\psi)\simeq1.26$, and used cutoff sequences with spacing $\Delta E_{\rm max}/m_\psi=1.25$. The oscillations visible in the lower panel occur on approximately this scale. The gray lines connect shifted sets of cutoffs with $\Delta E_{\rm max}/m_\psi=1.25$ and show that each set forms a smooth sequence, even though different shifts sample different positions relative to the discrete spectrum. The lowest of the sequences shown, with $E_{\rm max}/m_\psi = \cdots,\, 21.25,\, 22.5,\, 23.75$, is the one used for the scaling plots in Sec.~\ref{sec:convergence}. As shown there, this spacing gives clean convergence for the individual gaps. The differences between these shifted sequences are small relative to the gaps, but the binding energy is a much smaller difference of gaps, so they can be comparable to the binding energy itself. This motivates averaging over cutoff values spanning approximately one oscillation period, as done in Sec.~\ref{sec:BEs}.

\clearpage

\bibliographystyle{JHEP}
\bibliography{draft}

\end{document}